\documentclass[aps,prd,tightenlines,showpacs,nofootinbib,preprint,11pt]{revtex4}
\usepackage[english]{babel}
\usepackage{amsmath}  
\usepackage{amssymb} 
\usepackage{amsbsy}
\usepackage{amsfonts} 
\usepackage{graphicx}
\usepackage{epsfig}
\usepackage{mathrsfs}
\newcommand{\bec}{\begin{center}}
\newcommand{\eec}{\end{center}}

\def\?{?`}

\def\hil{\mathscr{H}}  
 
\newcommand{\beq}{\begin{equation}}
\newcommand{\eeq}{\end{equation}}
\newcommand{\beql}[1]{\begin{equation} \label{#1}}
\newcommand{\beqar}{\begin{eqnarray}}
\newcommand{\eeqar}{\end{eqnarray}}
\newcommand{\equref}[1]{Eq.~\eqref{#1}} 

\newcommand{\kets}[1]{\left\vert #1 \right\rangle}
\newcommand{\bras}[1]{\left\langle #1 \right\vert}
\newcommand{\ketsl}[1]{\left\vert #1 \right\rangle\!\!_L}
\newcommand{\brasl}[1]{{}_L\!\!\left\langle #1 \right\vert}
\newcommand{\ketsr}[1]{\left\vert #1 \right\rangle\!\!_R}
\newcommand{\brasr}[1]{{}_R\!\!\left\langle #1 \right\vert}
\newcommand{\bkt}[2]{\left\langle #1  |   #2\right\rangle}
\newcommand{\abs}[1]{\left\vert #1 \right\vert}
\newcommand{\ev}[1]{\langle #1 \rangle}
\def\omom{\bkt{0_M}{0_M}}
\def\bm{\bras{0_M}}
\def\km{\kets{0_M}}
\def\tr{\mathrm{Tr}} 
\def\ffi{\hat{\phi}} 
\def\ro{\hat{\rho}} 

\newcommand{\etal}{\hbox{\em et.al.}}
\newcommand{\wrt}{\mbox{w.r.t.}}

\newcommand{\rhs}{\hbox{r.h.s.}}
\newcommand{\cf}{\hbox{cf.}}
\numberwithin{equation}{section}

\def\th{\theta}
\begin{document}
\title{On the puzzle of Bremsstrahlung as described by coaccelerated observers}
\date{\today}
\author{I.~Pe\~na}\email{igor@nucleares.unam.mx}
\author{C.~Chryssomalakos}\email{chryss@nucleares.unam.mx}
\author{A.~Corichi}\email{corichi@nucleares.unam.mx}
\author{D.~Sudarsky}\email{sudarsky@nucleares.unam.mx}
\affiliation{Instituto de Ciencias Nucleares,  Universidad Nacional Aut\'onoma de M\'exico\\
Ciudad Universitaria,  A.P. 70-543, 04510, D.F., M\'exico}
\begin{abstract}
 We consider anew some puzzling aspects of the equivalence of the quantum field theoretical description of Bremsstrahlung 
from the inertial and accelerated observer's perspectives. More concretely,  we focus on the  seemingly
 paradoxical situation that arises when 
noting that the  radiating source is in thermal equilibrium  with the thermal state of the quantum field in the wedge in which it is
 located, and thus its presence  does not change there the state of the field, while it clearly does not affect the state of the field  on the opposite wedge.  How  then is the state of the quantum field on the future wedge changed, as it must in
 order to account for the changed energy momentum tensor there? This and related issues are carefully discussed.
 \end{abstract}
\maketitle
\section{Introduction}  
\label{sec:intro}

\begin{figure}[!b] \label{fig1}\bec
\includegraphics[scale=.65]{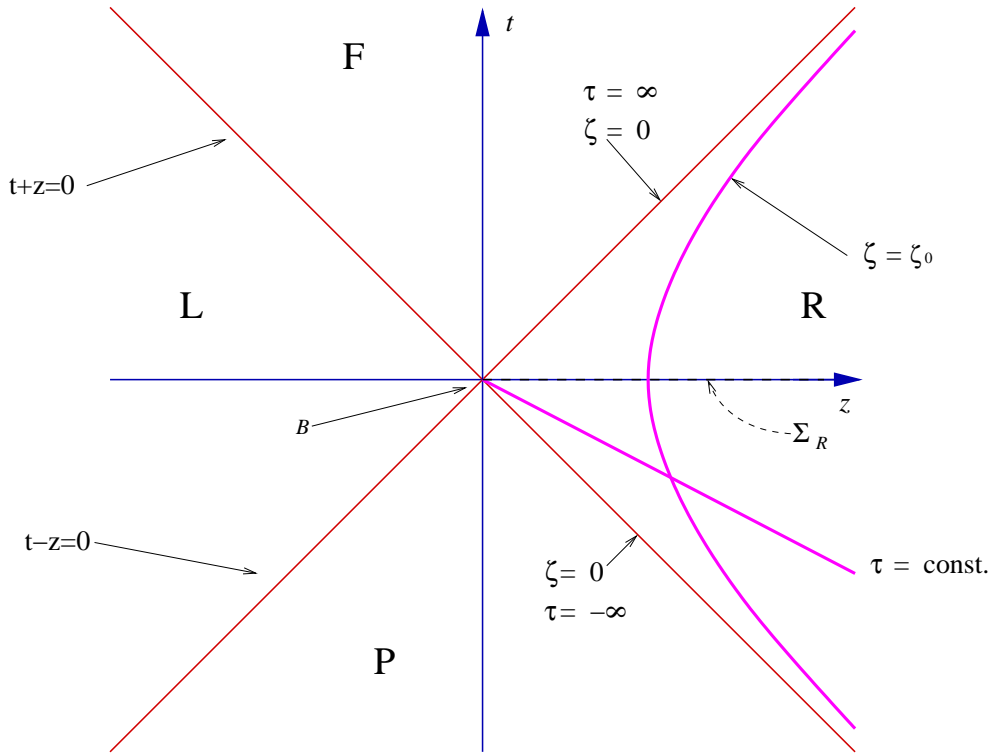}
\eec
\caption{The geometry of Rindler space.}\end{figure}

The topic of radiation by uniformly accelerated  charges has often been the source of much puzzlement and confusion, 
particularly when considered 
in the light of the equivalence principle. Much of the confusion has been removed by the realization that for 
observers coaccelerating  with  the charge there are regions of spacetime that are inaccessible. 
In effect, in the classical context, 
it has been shown that for an  electromagnetic charge in uniform acceleration, 
the classical radiation field,  as described by accelerating observers, is zero at every point  in the region that is accessible to them
(known as the Rindler wedge,  $R$, see Fig.~\ref{fig1})~\cite{boulware}.  This would then remove any apparent 
contradiction between the EP and the known Bremsstrahlung. For both the inertial and accelerated observers there is 
radiation but, for the accelerated ones, such radiation
lies beyond the regime where the static description is valid.

In the quantum version of this situation   the question is posed in terms 
of emission of photons rather than  the evaluation of radiation fields.
In fact, it was shown that the standard Bremsstrahlung when viewed from the point of view of the 
accelerated observers --a point of
view called Rindler quantization--  acquires a
very particular interpretation.
Actually, as will be explained in more detail below,   the coincidence  for the prediction of photon 
emission rates between the inertial and accelerated descriptions makes fundamental use of 
the Unruh effect~\cite{efectounruh}, which states that from an accelerated frame
comoving with the charge,  the standard Minkowskian vacuum state corresponds to  a thermal state. 
Moreover, it is well known that
for a detector uniformly accelerating  in the inertial vacuum the process for which the detector
absorbs a particle from the bath (a Rindler particle) as seen by a  comoving (accelerating) observer
is equivalent to the emission from the detector of a Minkowski  particle as seen by an inertial observer~\cite{waldunruh}.  
Then from the point of view of an inertial observer the   accelerating charge emits particles while, from an accelerated viewpoint,
the charge --which is static-- will emit \emph{and} absorb Rindler particles to and from the bath.   

The restriction of this effect to wedge $R$ has been analyzed 
in Ref.~\cite{bremstralung}, where it is shown  that  the emission rate 
of photons with fixed transverse momentum in the inertial frame coincides with the combined rate of emission and 
absorption of zero-energy Rindler photons with the same transverse momentum in the accelerated frame. 
Thus, this result gives a clear notion of the 
the physical equivalence between inertial and accelerated descriptions of Bremsstrahlung. However, and as is often the case in this field, the answer to one question brings in further 
puzzlement, and the need to answer a further one.

The calculation mentioned above,  makes  fundamental use of the so called zero energy Rindler modes.
The reason is  that for accelerated observers the charge is static and thus  it can only couple to 
modes of zero frequency with respect to Rindler time. In fact, due to the expression of the form $0 \times \infty$ that appears
in calculations involving zero energy particles in 
the analysis of  Ref.~\cite{bremstralung}, it is  required the introduction, for the purpose of 
regularization,  of a small frequency of oscillation $\th$ for the source, which
allows one to work with finite energy modes and which at the end is taken to 0.  
In that  work the authors considered also the question of whether or not an accelerated observer
sees any difference
in the thermal bath due to the emission and absorption of Rindler photons. They note that in the limit of 
zero energy the transition rate from an $n$-photon state to an $(n+1)$-photon state and the rate of the inverse 
process became equal, thus the accelerated charge leaves the thermal bath undisrupted. From this one can conclude that  \emph{the source
is in thermal equilibrium with the quantum field.}
Then, from the point of view of  an accelerated observer,  in $R$, there  will  be  no difference between the initial  state
of the field (the initial thermal state) and the state generated by the interaction with the accelerating charge  
(note the analogy with the description of the  situation in the classical context that we 
explained above).   Similarly, the state of the field in the second Rindler wedge will remain the initial thermal state, as
that region could not possibly be, in anyway, influenced by what is going on in a
causally disconnected region of spacetime.
We could think that this conflicts with the fact that EPR influences are allowed but this is misleading: when talking about the state in the $L$ Rindler wedge we mean the corresponding density matrix and this could not change due to actions taken on the right wedge  for otherwise there would be operators pertaining to the left wedge whose expectation values could be used  to determine  what has occured in the right Rindler wedge.
Nevertheless, it is clear that  in wedge $F$ (see Fig.~\ref{fig1}) there would be a detectable  change in the 
state of the field and, in particular, the expectation of the energy momentum tensor in  this region should be different from what it would have been 
if there was no interaction with the charge. In effect, this change could be computed  in an inertial  quantization of 
the field and would correspond to the final state containing  the  Minkowski photons emitted by the  the accelerated charge.
On the other hand, the quantum description of the state of the field  as seen by  the accelerated observer is given by the Unruh quantization 
scheme~\cite{efectounruh,librowald}. In this description, the  restriction to the $L$ and $R$ wedges of  both, the Minkowski vacuum state 
and the state resulting from the interaction with the charge  
is, in both cases, a thermal bath, and thus it seems neither could contain the information regarding what has changed in $F$.
The issue is then whether this situation can be analyzed in the language appropriate to accelerated observers 
and how would, in that case, the information about the changed
situation in $F$ be codified?

The fact that Unruh modes  can be expressed by  superpositions of 
specific  modes  that extend distributionally to the whole spacetime 
(known as \emph{boost modes}, see Refs.~\cite{ruskis,fullvsruskis}), 
opens the door to analysis of issues related to physical questions outside the double wedge,
and in particular in wedge $F$, carried out in the language appropriate for accelerated
observers. 
In this work, we embark on such analysis for the case of  a scalar field and an accelerated scalar source,  
and  analyze from the new perspective the change in the
expectation value of the energy-momentum tensor operator when evaluated at points in wedge $F$.

States in the Unruh quantization  
can be seen as states of a composite system where each of the components is 
the quantum field restricted to either wedges $L$ or $R$.  Then, having the density
matrix $\ro$ for a state, one can describe 
the physics, for example, in wedge $L$ ($R$)
by  tracing out in $\ro$ the right (left)  degrees of freedom. From this procedure
we obtain a density matrix $\ro^L$ ($\ro^R$) describing a state in wedge $L$ ($R$).
It is well known that when a field observable $\hat{A}$ is localized
in either $L$ or $R$,
then the expectation value of $\hat{A}$
is determined \emph{completely} by $\hat{\rho}^L$ or $\hat{\rho}^R$ respectively. 
On the other hand, when $\hat{A}$ is localized out of the double wedge, 
it is clear  that its  expectation value
would not, in general, be determined solely
by information encoded in $\hat{\rho}^L$ or $\hat{\rho}^R$.

There should exist some object containing this extra information 
carried  by the state which controls how do left 
and right parts \emph{combine} which  is the extra element  necessary
for describing completely the state in all of Minkowski spacetime. 
We identify this object as the \emph{entanglement  matrix}. In the 
case of interest we were concerned about the expectation value of the energy momentum tensor in the future wedge and it seemed natural to expect
its change to be encoded in the change of the entanglement matrix.
This was our initial assumption, 
and the work intended to see how
exactly is such information encoded in this case. As we will see this expectation was mistaken and
the answer in our case lies, surprisingly, in $\ro^R$.  
Precisely how does this matrix codifies this information will be elucidated through the rest of the manuscript.

Regarding the foundational basis of this work we must point out that the issue of \emph{extending} the Unruh quantization outside 
the double wedge is somewhat subtle. What
is very well established  in the literature is the restriction of global states to the double wedge --for
example, the  restriction of  states to the wedge $R$ is interpreted as the state seen by accelerated observers 
which have available only the wedge $R$-- but, as far as we know, the extension of a state in the Unruh quantization to the whole
MS  does not seem to be, at this time,  fully  investigated. 
One  important point concerning this question is  that   
Unruh modes 
are highly \emph{singular} at the asymptotes which are taken to coincide with the horizon of the
particular construction.  
Thus, the initial data from
which one builds up the (one particle) Hilbert space in the Unruh quantization has restrictions at these asymptotes
which  \emph{can influence} the solutions of the field equation in wedges $F$ and $P$ (this possibility has been pointed out
 in Ref.~\cite{fullvsruskis}, \cite{ruskis}). 
As we explained above, in this paper we extend 
the usage of the Unruh description of the field to address questions related to the physics in the wedge $F$.
We do this in the way that we consider is the most natural and we find results that are physically consistent and which coincide with what one would obtain working  in the standard Minkowski quantization of the field as we show explicitly in 
Appendix~\ref{app:inertial}.

The paper is organized as follows. In Sec.~\ref{sec:unruh} we review the main ideas of the Unruh quantization and specify our notation. In Sec.~\ref{sec:pos} we explain our strategy for analyzing the encoding of the information in the final state of the field. We propose a particular decomposition of the density operator of the state and 
use it to write 
an  expression for the change in the expectation value of $\hat{T}_{\mu\nu}$. Nevertheless, the particular form of the source representing an accelerating particle is needed in order to obtain an explicit expression for the final state and its density operator. In Sec.~\ref{sec:QR} we work out these ideas introducing a scalar accelerating pointlike source with a particular regulator and build up the S-matrix operator.
In Sect.~\ref{sec:eval} we make the final calculations in order to obtain the 
change in $\ev{\hat{T}_{\mu\nu}}$ and we evaluate it in wedges $R$ and $F$. From these results one can see how the information of the physical change in $F$ is encoded.
In Sec.~\ref{sec:ex} we analyze the case of two different sources accelerating in wedges $L$ and $R$ in order to get more insight of the behaviour of the density operator of the respective final state. Finally, we end with some discussions  and the 
interpretation of the results in Sec.~\ref{sec:dis}.
In order to carry out these steps it  is important to have explicit expressions of the spacetime behaviour of the Unruh modes, which we present in Appendix~\ref{sec:repre}. The calculations leading to our final results are, though straightforward, very long, in Appendix~\ref{app:scnd} we give an explicit derivation of the main formula in this work.
In Appendix~\ref{app:inertial} we make the same calculations as in Sec.~\ref{sec:eval} for the change in wedge $F$  but using, instead, the standard plane wave representation of the field and show that both results coincide.

To eliminate unnecessary notation we shall work in two dimensional Minkowski space, this will allow us to be clearer
without losing physical insight. As a matter of fact, working with a scalar field in $2D$ with $m\neq 0$ is operationally equivalent to work
with the same field in 4D and fixed traverse momentum 
$k^2_\perp = k_x^2 + k_y^2 $, with the identification $m^2 \to k_\perp^2  + m^2$. 
In all this work we shall work in units in which $c=\hbar = 1$ and signature $-\,+$.

\section{Unruh quantization}
\label{sec:unruh}

All the comoving observers to a particle moving with uniform proper acceleration  $a=(a_\mu a^\mu )^{1/2}$ 
have world lines of the form
\beql{ccor}
t=\zeta  \sinh (a \tau) \qquad z=\zeta \cosh (a \tau) \, ,
\eeq
where  $0 < \zeta < \infty$ and   $\tau$ is the proper time of the observer,
$ -\infty<\tau<\infty$. \equref{ccor} can be used to give coordinates
$\tau,\zeta$ to wedge $R$, which is known as Rindler spacetime. In these coordinates, 
the  Minkowski metric becomes%
\beql{metR}
ds^2 = - \zeta^2 d\tau^2 + d\zeta^2 .
\eeq 

As can be seen from Eq.\eqref{metR}, wedge $R$ is a static, globally hyperbolic spacetime so one can use
standard methods of frequency splitting~\cite{librowald,kay,ashtekar} for quantizing the field in that region. As a first step we shall construct
directly  the one particle Hilbert space for a Klein Gordon field with mass $m$ in wedge $R$, using as timelike Killing field $\tau^a$; this
is called the Fulling-Rindler quantization \cite{fulling}.
Consider the massive Klein Gordon (KG) field equation 
\beql{kg}
\big( \Box - m^2 \big) \phi = 0
\eeq
restricted to wedge  $R$
and take solutions which are \emph{positive frequency} \mbox{w.r.t.} the Rindler time $\tau$ 
(in this case $(\partial/\partial \tau)^\mu = b^\mu$ is the time translation generator, \cf{} \equref{ba})
and  which vanish asymptotically. 
These are superpositions of the following set of modes
\beql{fullmod}
\psi_\omega (\xi) = \frac{\sqrt{\sinh{\pi\omega}}}{\pi} e^{-i\omega \tau} K_{i\omega} (m\zeta) ,\qquad \omega>0,
\eeq
where $\xi = (\tau,\zeta)$, and $K_{i\omega} (x)$ is the modified Bessel function of the third kind or
Macdonald function. The normalization factor is chosen such that these modes are $\delta$ normalized \mbox{w.r.t.} 
the KG product: 
\beql{kgnormm}
\bkt{\psi_\omega}{\psi_{\omega'}}_{KG} = \frac{i}{2} \int_{\Sigma_R} \big( \psi_\omega^* \nabla_\mu \psi_{\omega'} 
- \psi_{\omega'} \nabla_\mu \psi_\omega^*  \big) d\Sigma^\mu = \delta ( \omega - \omega' )
\eeq
where $\Sigma_R$ is a Cauchy surface for Rindler spacetime.
Functions $K_{i\omega} (m\zeta)$ have an essential singularity (it oscillates ``infinitely'') in the 
limit  $\zeta\to 0$~\cite{watson}, 
so the $\psi_\omega (\xi)$ are not defined at  the horizon ($\tau \to \pm \infty$ and $\zeta \to 0$) when $\omega\neq 0$.
Modes given by  Eq.~\eqref{fullmod}, called Fulling modes,   and their conjugates $\psi_\omega^* (\xi)$ form a complete set of the space of solutions 
to Eq.\eqref{kg} in Rindler spacetime.  One constructs the one particle Hilbert 
space $\hil_R$ for a quantization of the KG field in $R$ by Cauchy completing the space spanned by the $\psi_\omega (\xi)$ 
(positive frequencies) in the inner product given by Eq.~\eqref{kgnormm}   (a rigorous construction of this space is given in~\cite{kaydouble}).
The space of states of the field for this quantization  is the Fock space of $\hil_R$, $\mathscr{F}(\hil_R)$.
The field operator in this quantization takes the form
\beql{firr}
\ffi_R (\xi) = \int_0^\infty \left( \psi_\omega (\xi) \hat{{r}}_\omega + \psi_\omega^*  (\xi) \hat{{r}}^\dagger_\omega\right)\,d\omega ,
\eeq
where the creation and annihilation operators, $\hat{{r}}^\dagger_\omega$ and  $\hat{{r}}_\omega $, satisfy 
canonical commutation relations. 
Repeating the Fulling-Rindler  quantization procedure in wedge $L$ with the $L$-time translation generator
given by $-b^\mu$ (see Eq.~\eqref{ba}) and defining the KG product on a Cauchy surface $\Sigma_L$ of $L$,  one obtains the one particle Hilbert space
for the KG field in $L$, $\hil_L$, and a \emph{left} field operator analogous to \equref{firr}.

Wedges $L$ and $R$, are causally disconnected  and 
thus neither $\mathscr{F}(\hil_L)$
nor $\mathscr{F}(\hil_R)$ can be used to represent the field algebra in Minkowski spacetime, as the Fock space of the standard plane 
wave quantization
does. However, one can consider a second quantum field construction in all of Minkowski spacetime  with one particle Hilbert space
given by
\beql{hilU}
\hil_U = \hil_L \oplus \hil_R ,
\eeq
and the  space of states given by~\cite{librowald}
\beql{focku}
\mathscr{F} (\hil_{U})=\mathscr{F}(\hil_{R})\otimes\mathscr{F}(\hil_{L}) \, .
\eeq
The quantum field construction defined by \equref{hilU} is called \emph{Unruh quantization}.
The field operator in this quantum construction takes the form
\beql{phiu}
\ffi (x) = \ffi_L (x) \otimes \hat{1}_R + \hat{1}_L \otimes \ffi_R (x) .
\eeq
Whenever there is no confussion, we denote by  $\ffi_{L}(x)$ the operator
$\ffi_L (x) \otimes \hat{1}_R$ and the analogous for $\ffi_R$.  Clearly,  these operators
commute,
\beq
\big[\ffi_L (x),\ffi_R (x')\big] = 0 ,
\eeq
reflecting the fact that  the regions  $L$ and $R$ are causally disconnected.

Constructed in this way, it may seem that in the Unruh quantization the field operator is made up of modes
which have support only in the double wedge $L\cup R$ and thus that cannot describe the physics outside this
region.  Nevertheless, recall that the one particle
Hilbert spaces $\hil_L$ and $\hil_R$ are made up of positive  frequency 
modes which have initial data on either $\Sigma_L$ or $\Sigma_R$. These two latter Cauchy surfaces
can be seen as the restriction of some Minkowski spacetime Cauchy surface $\Sigma$ to $L$ or $R$ respectively, then
the initial data of modes in the left or right quantizations of the field  
define unique solutions of
the KG equation in all of Minkowski spacetime. In fact, 
as we explain in the next subsection and in  Appendix~\ref{sec:repre}, there exist 
superpositions of plane waves, called Unruh modes, 
which coincide with Fulling modes  when restricted to either wedge $L$ or
$R$ and that are zero when restricted to the opposite wedge~\cite{ruskis}. The elements of $\hil_U$ describe modes of the 
field defined distributionally in the whole of Minkowski spacetime. See the last paragraph of the next subsection for 
further discussion.

\subsection{Boost modes and Unruh modes}
\label{subsec:bmum}

The original quantization approach used by Unruh \cite{efectounruh}  
 does not give explicitly the functional form of Unruh  modes since he works only with their restrictions to the horizons. However, for the 
purposes of this work we do need the functional form in all of MS of these set of  modes, which are linear combinations
of \emph{boost modes}. We will  introduce  both sets of modes in this section following Ref.~\cite{ruskis}. 

Consider the space of global  classical  solutions to Eq.~\eqref{kg}. One can look for solutions in this space which 
have positive frequency \mbox{w.r.t.} the boost parameter $\tau$
in wedges $L$ and $R$.  This class of solutions, 
\emph{Minkowski Bessel Modes} were originally presented and studied  by Gerlach~\cite{gerlach}. Narozhny~\etal{}~\cite{ruskis}
 and Fulling and Unruh~\cite{fullvsruskis} call them \emph{boost modes} (BM), name that we also prefer.
%
Take an (unnormalized)  plane wave solution to Eq.~\eqref{kg} of the form
\beql{pw}
P^\pm_p (x) = (2\pi)^{-1/2}  \; e^{\mp  i ( \omega_p t -  pz)} ,
\eeq
where $x=(t,z)$,  $\omega_p = \sqrt{p^2 + m^2} > 0$ and the upper (lower) sign corresponds to positive (negative) frequency
{\wrt} inertial time. Changing the variable $p$ to the \emph{rapidity}
${\th}$:
\beq
m \sinh (\th) = p ,\qquad m\cosh (\th) = \omega_p ,\qquad -\infty < \th < \infty ,
\eeq 
then one can define  boost modes as the 
following superposition of plane waves~\cite{ruskis}
\beql{bm}\begin{split}
B^\pm_\omega (x) & = \frac{1}{2^{1/2}\sqrt{2\pi}}  \int_{-\infty}^\infty P^\pm_\th (x) e^{-i\omega \th} \, d\th \\
& = \frac{1}{2^{3/2}\pi} \int_{-\infty}^{\infty} e^{\mp im (  t\cosh \th  -  z \sinh  \th)}  e^{-i\omega \th}\, d\th ,
\end{split}\eeq
where $-\infty < \omega < \infty$. 

Note that in general,  \equref{bm} cannot be interpreted as the definition of a function.
In particular, 
it is divergent at  the origin, $(t=0,z=0)$ and thus,  cannot stand as a 
global solution to the KG equation if it is considered as a function. 
However, one can avoid these problems if one considers these quantities as
distributions, thus requiring the smearing with suitable test functions.
Therefore, we will consider this set of modes, as well as
Unruh modes (\cf{} Eqs.~\eqref{um}), as distributions
when constructing the quantum field. Note that this is a consistent procedure 
 as the field $\ffi(x)$ has, by itself, a distributional character. 
For the purposes of this work,  
we will consider only test functions of compact support in $M$. 

Formally, boost modes  are orthogonal in the KG inner product (Eq.~\eqref{kgnormm} over $\Sigma_M$),
\beq
\bkt{B_\omega^\pm}{B_\mu^\pm}_{KG}=\pm\delta(\omega-\mu),
\eeq
and they   are eigenfunctions of the \emph{boost operator}~\cite{gerlach}
(inside the four wedges of MS),
\beql{bop}
\mathbf{B} B^\pm_\omega = - i \omega  
B^\pm_\omega \qquad \text{where} \qquad  \mathbf{B} = t\frac{\partial}{\partial z} + z \frac{\partial}{\partial t} = 
\frac{\partial}{\partial \tau},
\eeq
and thus, the modes  $B^\pm_\omega$ are \emph{positive (negative)} frequency KG solutions {\wrt} the boost parameter $\tau$
in wedges $L$ and $R$ whenever $\omega > 0$ ($\omega < 0$). 
Note that in dividing the modes into positive and negative frequency modes one is dropping
out the $\omega =0$ mode  which could, in principle,  be a source of trouble~\cite{ruskis}.

From now on we will only use boost modes which are positive frequency {\wrt} inertial time $t$, $B_\omega^+$, 
and will drop the superscript: $B_\omega \equiv B_\omega^+$. 
One can use this set of modes as a \emph{basis} of
the positive inertial frequency solution space of eq.~\eqref{kg}. Actually, one can give a   
quantization of the KG field unitary equivalent to the standard positive frequency plane waves one~\cite{ruskis}:
\beql{fibm}
\hat{\phi} (x) = \int_{-\infty}^{\infty} d\omega \; \big( B_\omega (x) \hat{b}_\omega +
B_\omega^* (x) \hat{b}_\omega^\dagger \big),
\eeq
where $[\hat{b}_\omega , \hat{b}^\dagger_{\omega'}]=\delta (\omega - \omega')$. Because the transformation is 
unitary these quantum field descriptions share the same vacuum
\beql{misvac}
\hat{b}_\omega \km = \hat{a}_p \km = 0.
\eeq
Unruh's idea~\cite{efectounruh}  for giving a field quantization associated to accelerated observers consists in constructing
a set of  modes $R_\omega$ and  $L_\omega$ from  combinations of boost modes  and their conjugates such that they
are positive frequency {\wrt} the boost parameter (accelerated time) in
wedges $R$ and $L$  respectively
 and zero on the opposite wedge. These modes are defined by
\begin{subequations}\label{um}\begin{align} 
R_\omega (x) & = \frac{1}{\sqrt{2\sinh(\pi\omega)}} \left[ e^{\pi\omega/2}B_\omega (x)  - e^{-\pi\omega/2} B^*_{-\omega} (x) \right],
\label{Rw} \\
L_\omega  (x) & = \frac{1}{\sqrt{2\sinh(\pi\omega)}} \left[ e^{\pi\omega/2}B_{-\omega} (x)  - e^{-\pi\omega/2} B^*_\omega (x) \right],
\label{Lw}
\end{align}\end{subequations}
where $\omega > 0$. Note that, since they are  defined in terms of boost modes, these definitions are global. These equations can be inverted:
\begin{subequations}\label{mu}\begin{align}
B_\omega (x) &= \frac{1}{\sqrt{2\sinh (\pi\omega)}} \big[ e^{\pi\omega/2}R_\omega (x) + e^{-\pi\omega/2} L_{\omega}^* (x) \big],
\label{Bw}\\
B_{-\omega} (x) &= \frac{1}{\sqrt{2\sinh (\pi\omega)}} \big[ e^{\pi\omega/2}L_\omega (x) + e^{-\pi\omega/2} R_{\omega}^* (x) \big].
\label{Bmw}
\end{align}\end{subequations}

It can be seen that, when restricted to $R$,  Unruh modes $R_\omega$ coincide with  Fulling modes, Eq.~\eqref{fullmod},
and the analogous situation for $L_\omega$ in $L$~\cite{gerlach} (see Appendix~\ref{sec:repre}) and, therefore,  initial data of Unruh modes coincides with that of Fulling modes. It is known that solutions of the field equation can be represented by their initial data \cite{kay} and thus Cauchy completing 
the space spanned by the $L_\omega$ and 
$R_\omega$ modes one would obtain, respectively, $\hil_R$ and $\hil_L$.  

The field operator in the Unruh quantization  takes the form of \equref{phiu} with $\ffi_L (x)$, $\ffi_R (x)$
expressed in terms of Unruh modes:
\beql{fiun}
\hat{\phi} (x)  = \ffi_L (x) + \ffi_R (x) = \int_{0}^{\infty} d\omega \big( R_\omega (x) \hat{1}_L\otimes\hat{r}_\omega 
+ L_\omega (x) \hat{l}_\omega\otimes\hat{1}_R
+ \mathrm{H.C.} \big),
\eeq
where $\hat{r}_\omega$ and $\hat{l}_\omega$ are annihilation operators in $\mathscr{F}(\hil_R)$ and $\mathscr{F}(\hil_L)$
respectively. 

We note that recently there has been some controversy about the Unruh quantization.
Narozhny \etal{}~\cite{ruskis} claim that the \emph{Unruh
quantization is not a valid quantization scheme for all of MS} (as the boost modes quantization is, see Eq.~\eqref{fibm}).
To support this they argue, in particular,  that the expansion of $\hat{\phi}$ in Unruh modes, Eq.~\eqref{fiun}, does not exhaust all
the degrees of freedom of the quantum field in Minkowski space. 
This is so, they say,  because when ``evaluating at the origin'', boost modes have a \emph{singularity} when $\omega = 0$.
The authors of Ref.~\cite{ruskis} logic is that, consequently, in the transition 
from Eq.~\eqref{fibm} to  Eq.~\eqref{fiun} ---using Eqs.~\eqref{mu}---
one should take the Cauchy principal value of the integral at the 
origin, which excludes the $\omega=0$ mode. They assert that without this mode, the remaining set of
boost modes loses the property of being complete and then  lacks the possibility of spanning every (one particle)
state of the field. (Ref.~\cite{ruskis}, p.~025004-12)
On the other hand, Fulling and Unruh~\cite{fullvsruskis} have argued against  this claim. They state that since the mode expansion 
is an integral (Lebesgue measure) and thus one mode is of \emph{zero measure}, the omission of the $\omega=0$ mode is quite  
harmless in the mode expansion  of the field and the Unruh quantization is valid for expressing the restrictions to $L\cup R$ of global MS states. 
In this work we adhere to the position of Fulling and Unruh  for the following  reasons.  As we have said above, boost modes should be considered
as \emph{distributions} and thus, evaluating them at  one single point has no meaning.
Moreover, we recall that the field itself evaluated at one point has no meaning either, the
quantum field is a distributional object and only its convolution with a test function 
is defined. 
  
In their reply, Fulling and Unruh forcefully argue  that the Unruh quantization scheme is
valid on the double wedge $L\cup R$ but, however, they indicated that they are not fully confident   on the possibility of \emph{extending} an  Unruh state
to all of MS (\cite{fullvsruskis} p.048701-2). Let's be more precise on this issue.
The initial data for the Unruh quantization  consists of smooth functions of compact support on both Cauchy surfaces
$\Sigma_R$ and $\Sigma_L$. For definiteness, take $\Sigma_R = \{(0,z)\vert z>0\}$ and $\Sigma_L = \{(0,z)\vert z<0\}$.
For these Cauchy surfaces, functions of compact support would be zero at the origin $(0,0)$. Nevertheless, a full MS quantization
should consider the set of all $C_0^\infty$ initial data over the Cauchy surface 
\beq
\Sigma_M = \Sigma_L \cup \Sigma_R \cup \{ 0 \},
\eeq
which includes,  of course, functions which are not zero at the origin. In this respect, Fulling and Unruh~\cite{fullvsruskis} 
note that ``The treatment of initial data at the origin is mathematically subtle, and data at that point may influence
the solution of the field equation in regions $F$ and $P$''.  
This issue is quite relevant for our work. Although we are not giving any formal proof of the fact that the 
Unruh quantization \emph{can be extended} to all of MS, we show that at least for the particular case 
we study, the  Unruh quantization
provides the same 
physical results  as the standard flat space  quantization. 
This we take as an indication that both quantum descriptions are generally equivalent.

\subsection{Minkowski vacuum, general states and $L$-$R$  correlations}
To find a basis of $\mathscr{F} (\hil_{U})$, \equref{focku}, we need a  basis for $\mathscr{F} (\hil_{R})$ and
$\mathscr{F} (\hil_{L})$.
For  $\mathscr{F}(\hil_{R}) $  we choose an orthonormal basis  whose elements are  $R$-states $\kets{J}_R$ which have 
a definite number of Rindler particles $n_J$. 
Let  $J_{\omega_m}$ be the number of particles in this state whose frequencies are centered in the particular mode $\omega_m$, 
$m=0,1, \dots$.
Then, the  state $\kets{J}_R$ is defined by the set 
\beql{defJ}
J=\big\{J_{\omega_0}, J_{\omega_1}, \dots , J_{\omega_m} , \dots       \big\} \; , \qquad \sum_{m=0}^\infty J_{\omega_m} = n_J \, , 
\eeq
Note that only a finite number of the  $J_{\omega_m}$ are $\neq 0$.
State  $\ketsr{J}$ can be build up from the Rindler vacuum in the following manner
\beq
\ketsr{J} = \ketsr{J_{\omega_0} \, J_{\omega_1} \cdots} = N_J \, (\hat{r}_{\omega_0}^\dagger)^{J_{\omega_0}} 
\cdots (\hat{r}_{\omega_m}^\dagger)^{J_{\omega_m}}\cdots \ketsr{0},
\eeq
where $N_J$ is a normalization factor such that $\bkt{J}{J}=1$, and $\hat{r}_{\omega_m}^\dagger$ creates a Rindler
particle with frequency centered at $\omega_m$.
 State $\ketsr{J}$ has Rindler energy
\beql{e(j)}
E(J) = \sum_{m=0}^\infty \omega_m J_{\omega_m}.
\eeq
This is  the state's energy 
associated with the boost Killing field.
Given two different elements $\ketsr{J}$, $\ketsr{K}$ of this basis
we have that
\beql{ladelta}
{}_R\!\bkt{J}{K}_{\! R} =  \delta (J,K) \equiv \delta_{J_{\omega_0},K_{\omega_0}} \cdots \delta_{J_{\omega_m} ,K_{\omega_m}} \cdots,
\eeq
where $\delta_{J_{\omega_m} , K_{\omega_m}}$ are  Kronecker deltas.
One can obtain a basis of $\mathscr{F}(\hil_L)$ in an analogous way. The set of all $ \kets{J}_L \otimes \kets{K}_R$ where
$J$ and $K$ are of the form of Eq.~\eqref{defJ} is a basis of $\mathscr{F} (\hil_U )$.
Any state $\kets{g} \in \mathscr{F}( \hil_U )$ can be casted in terms of this basis, and it is defined by  
a particular  function $G(J,K)$:
\beql{estgen}
\kets{g} = \sum_{J,K} G(J,K) \ketsl{J}\ketsr{K}.
\eeq
Here, the sums run over the space of all possible \emph{particle distributions} $J$ and $K$, that is, they 
are of the form
\beql{suma}
\sum_J = \prod_{m=0}^\infty \sum_{K_{\omega_m}=0}^\infty  .
\eeq
Entangled  states in the Unruh quantization (those which  cannot be expressed in the form $\ketsl{g_1} \otimes \ketsr{g_2}$) 
have non trivial correlations between left and right states.
The inertial Minkowski vacuum, $\km$, is an entangled state in the Unruh quantization:
\beql{0mcorr}
\km = \sum_K e^{-\pi E(K)} \ketsl{K} \ketsr{K} = \sum_{J,K} e^{-\pi E(K)} \delta(J,K) \ketsl{J}\ketsr{K} .
\eeq
Given that  $\km$
is an entangled state, when restricted  
to the $R$ wedge, it  fails to be a 
pure state~\cite{rovelli} and its description is in terms of a density matrix. The norm of the Minkowski vacuum
is now given by
\beq\label{0mnorm}
\bkt{0_M}{0_M} =  \prod_{m=0}^\infty \frac{1}{1-e^{-2\pi\omega_{m}}} .
\eeq
The fact that this well behaved state in the plane wave quantization has 
an infinite norm in the Unruh quantization is a consequence of the non unitarily equivalence between these
quantizations.   
Recall that  different, even not unitarily equivalent,  representations of the field algebra are ``physically equivalent'' in the sense of Fell's
theorem  (see Ref.~\cite{librowald}) and thus one expects to obtain the same physical information from the computation of expectation values  in either the inertial or the Unruh scheme, provided that the Unruh quantization is a faithful representation of the field algebra\footnote{We thank Hanno Sahlmann for pointing this out.}. We can say that our results point in this direction since, as we shall see in Sec.~\ref{sec:eval} and Appendix~\ref{app:inertial}, we obtain, in both quantization schemes, the same physical result. 

\section{Posing the problem: the matrix of entanglement}
\label{sec:pos}
In  this work we shall be concerned with the difference between the initial vacuum state and the final state which results from  the initial state and the interaction of the classical scalar source in uniform acceleration.
At the same time we are interested with the restriction of these sates to the different regions
and thus we shall employ the language of density matrices as indicated by the posing of the questions raised in the introduction.

The trajectory of this source is a branch of an hyperbola lying  in one Rindler wedge of spacetime which we choose it to be $R$. The final  state of such interaction  can be obtained  perturbatively by the application of the respective   S-matrix
operator  to the inertial vacuum state (this operator will be constructed in detail in  Sect.~\ref{sec:QR}):
\beql{fSo}
\kets{f}=\hat{S}\km .
\eeq
Recall from Sec.~\ref{sec:unruh} that the $L_\omega$ modes of the field
are zero when evaluated in wedge $R$ and thus the scalar source can only excite $R$ modes of the field. Then, for this case the $\hat{S}$ operator takes the form
\beql{s1sr}
\hat{S} = \hat{1}_L \otimes \hat{S}_R.
\eeq
On the other hand, expressed in the Unruh quantization scheme,  state $\kets{f}$ takes  the form 
\beq
\kets{f} = \sum_{J,K} F(J,K) \ketsl{J} \ketsr{K}.
\eeq
All the information of the state $\kets{f}$ is encoded in the function $F(J,K)$, and hence
one should be able to read from
the change in this function, $\Delta F = F(J,K)-F_{vac} (J,K)$, the  
change in the energy momentum tensor due to the effects of the source.
In particular, this change should, at the same time, codify the fact that no change 
is produced inside the wedge $R$ and that a dramatic change occurs in wedge $F$. 
Nevertheless,  this information is more clearly 
encoded in the change in the density matrix of the state,
$\Delta \ro = \ro_f - \ro_{vac}$    since it can be splitted directly into
left, right and \emph{entangled} contributions as we will explain
in the following.
This is a second reason for using the language of density matrices.

We already know that, when considered as a state of the composite system
$\mathscr{F}_L \otimes \mathscr{F}_R$,  the inertial vacuum state is an entangled 
state (see \cite{rovelli} for an intuitive explantion of this), and thus one  expects that also  state $\kets{f}$ will  be 
entangled. 
In this sense, since both states are pure, their density operators cannot be written in the form \cite{horodeckis1}
\beql{peres}
\ro^{\,\prime}=\ro^{\prime\,L} \otimes \ro^{\prime\,R} ,
\eeq
where $\ro^{\prime\,L}$ and $ \ro^{\prime\,R}$ represent the respective partial density matrices  defined by
\beq
\ro^{L,R} \equiv \tr_{L,R} \; \ro .
\eeq
Nevertheless,
one can introduce a \emph{traceless operator}  $\ro^e$ which encodes all the 
information of the entanglement of the state.
We propose that the density operators  for these  states can be written as
\beql{ro+roe}
\ro = \ro^L \otimes \ro^R + \ro^e  .
\eeq
In the Unruh quantization, the $\ro^e$ operator, which we will 
call the \emph{matrix of entanglement},
 encodes the information 
of the correlation between left and right components of the state. 
When computing expectations of operators localized in either wedges $L$ or $R$
the matrix of entanglement plays no role, in fact, 
it can be shown that for any operator of the form  $\hat{A}_L\otimes\hat{1}_R$  and a state described by \equref{ro+roe} one has that
\beql{olrf}
\tr ( \hat{A}_L \otimes \hat{1}_R \:  \ro ) =  \tr ( \hat{A}_L \, \ro^L) ,
\eeq
and thus $\tr ( \hat{A}_L \otimes \hat{1}_R \,\ro^e ) =0$
(and the analogous if the operator has the form $\hat{1}_L \otimes\hat{A}_R$).  The information encoded in $\ro^e$ can only be retrieved 
when computing expectations of operators with $L$ and $R$  components and thus, for the case of observables (made up of field operators) this information is only present when one evaluates the expectation values in wedge $F$. 
For example, in the computations of \cite{bremstralung} the information encoded in $\ro^e$ does not enter in the results, which in principle are \emph{incompatible} with the
change in $\ev{\hat{T}_{\mu\nu}}$ in wedge $F$. 
It is therefore that we expected that the information about this change  would be encoded  in $\ro^e$.

We can write the inertial vacuum $\km$ 
in the fashion of \equref{ro+roe}:
\beq
\ro_{vac}     = \ro_{vac}^L \otimes \ro_{vac}^R + \ro_{vac}^e \, .   \label{rovace} 
\eeq
From \equref{0mcorr} it follows that 
\beql{rovac}
\hat{\rho}_{vac} = \km\bm = 
Z\sum_{JK} E_J E_K  \ketsl{K}\ketsr{K} \; 
\brasl{J}\brasr{J} ,
\eeq
where we have defined
\beql{EJ}
E_J \equiv e^{-\pi E(J)}
\eeq
(see \equref{e(j)}), and we have introduced the normalization factor $Z$ defined
by
\beql{Zeta}
Z^{-1}\equiv \sum_J E_J^2
\eeq
in order to have $\tr(\ro_{vac})=1$.
Taking the partial $R$ and $L$ traces in \equref{rovac} we obtain
respectively
\beql{rorrol}
\ro_{vac}^L = Z\sum_J E_J^2 \ketsl{J}\brasl{J}  ,
\qquad
\ro_{vac}^R= Z\sum_J E_J^2 \ketsr{J}\brasr{J}  .
\eeq
Using Eqs.~\eqref{rorrol} and \eqref{rovac} we can write
\beq\begin{split}
\ro_{vac}^e & = \ro_{vac} - \ro_{vac}^L \otimes \ro_{vac}^R\\
& = Z\sum_{J,K} E_J E_K \Big( \ketsl{J}\ketsr{J} \brasl{K}\brasr{K}
- Z E_J E_K \ketsl{J}\ketsr{K} \brasl{J}\brasr{K}\Big).
\end{split}\eeq

For the density matrix for state $\kets{f}$ we have 
\beq
\ro_f  =  \ro_f^L \otimes \ro_f^R + \ro_f^e \label{rofe} .
\eeq
Now we turn back to our own specific case and concentrate particularly in the change
in the density matrices induced by the interaction of the field with the source.
Exploiting the fact that for the accelerated source the $\hat{S}$ matrix is an $R$ operator
one finds directly from \equref{fSo} and \equref{s1sr} that
\beql{SS}
\ro_f = \ro_{vac}^L \otimes \hat{S}_R \,\ro_{vac}^R \, \hat{S}_R^\dagger +
\hat{S} \,\ro_{vac}^e \,\hat{S}^\dagger .
\eeq
From this last equation we can identify
\beql{rors}
\ro_f^L = \ro_{vac}^L \qquad \ro_f^R = \hat{S}_R \,\ro_{vac}^R \,\hat{S}_R^\dagger
\eeq
and 
\beql{roefe}
\ro_f^e = \hat{S} \,\ro_{vac}^e \,\hat{S}^\dagger .
\eeq
We want now to express the change 
in the state of the field induced by the interaction in terms of the change of the density matrix.
Note  that when restricted to $L$  this change   is  $\delta \ro_f^L =\ro_f^L - \ro_{vac}^L=0$. 
The changes 
$\delta \ro_f^R = \ro_f^R - \ro_{vac}^R$ and 
$\delta \ro_f^e = \ro_f^e - \ro_{vac}^e$ deserve special attention since they  depend
on the operator $\hat{S}$, note that they are traceless.

The total change in the density matrix can be written as 
\beql{deltaro}\begin{split}
\delta \ro 
& = \ro_{vac}^L \otimes \delta \ro_f^R +
\delta \ro_{f}^L \otimes  \ro_{vac}^R + \delta \ro_{f}^L \otimes \delta \ro_{f}^R
+ \delta \ro^e\\
& = \ro_{vac}^L \otimes \delta \ro_f^R
+ \delta \ro^e ,
\end{split}\eeq
where $\delta\ro^e \equiv \ro_f^e - \ro_{vac}^e$ and we have used $\delta\ro_f^L = 0$. 
\equref{deltaro} thus reflects in a clear language what we know about the change in the quantum field.
As we mentioned in Sect.~\ref{sec:intro}, there is theoretical evidence that when one restricts state
$\kets{f}$  to wedge $R$ one would obtain  the same thermal bath as that of the inertial vacuum  \cite{bremstralung}, that is, in this wedge there is no effective change in the state. A priori one may think that this  means that
$\delta\ro_f^R =0$.  However, the state should have changed in order to account for the radiation emitted by the source, which should  be measurable in $F$. 
 Thus one might conclude 
 that all the information of the physical change in the state is encoded in the change in the matrix of entanglement, $\delta\ro^e$,  which can only be retrieved in wedge $F$. 
To be more specific, consider an operator of the form  $\hat{A}=  \hat{1}_L \otimes \hat{A}_R$
on $\mathscr{F}_L \otimes \mathscr{F}_R$. From  \equref{deltaro} one has that
the change in its expectation is given by
\beql{check}
\delta\tr(\hat{A}\ro) \equiv \tr (\hat{A} \, \ro_f ) - \tr ( \hat{A} \, \ro_{vac} ) =
\tr ( \hat{A}_R \;  \delta\ro_f^R ) .
\eeq
Note that the operator $\hat{A}_R$ can be localized \emph{anywhere} 
in Minkowski spacetime
as, for example, the operator $\hat{1} \otimes \ffi_R (x)$, which is not identically zero outside wedge $L$.
However, if the operator $\hat{A}$ has non trivial $L$ and $R$ components
then the change is given by
\beq
\delta\tr(\hat{A}\ro) =\tr ( \hat{A} \; \ro_{vac}^L \otimes \delta\ro_f^R ) + \tr (\hat{A} \; \delta \ro^e) ,
\eeq
that is, in this case the change in the expectation may come from both $ \delta\ro_f^R $ and
$\delta \ro^e$ contributions.

In order to understand the physical change in the state we shall evaluate the change in the expectation of the energy-momentum tensor operator $\hat{T}_{\mu\nu}$.
Recall that in Minkowski spacetime 
$\ev{\hat{T}_{\mu\nu} (x)}$ 
is defined in the point-splitting description as a coincidence limit~\cite{librowald}: 
\beql{tmn}
\ev{\hat{T}_{\mu\nu} (x)}_{f} = \lim_{x'\to x} t_{\mu\nu\,'} F(x,x') ,
\eeq
where 
\beql{Fxx}
F(x,x')=\bras{f}\hat{\phi}(x)\hat{\phi}(x')\kets{f}-\bm\hat{\phi}(x)\hat{\phi}(x')\km
\eeq
and $t_{\mu\nu'}$ is the differential operator 
\beq
t_{\mu\nu'} = \nabla_\mu \nabla_{\nu'} - \tfrac{1}{2} g_{\mu\nu}
(\nabla_\sigma \nabla^{\sigma'} + m^2 )  .
\eeq
Thus, all one needs to compute the expectation of $\hat{T}_{\mu\nu}$ is
the change in the two point function $\ev{\ffi(x)\ffi(x')}$. 
We have  from \equref{fiun} that 
\beql{fifi}
\ffi (x) \ffi (x') = \ffi_L (x) \ffi_L (x')+\ffi_L (x) \ffi_R (x')
+\ffi_R (x) \ffi_L (x')+\ffi_R (x) \ffi_R (x')  .
\eeq
To simplify the notation, let us define
\beq
\ffi \equiv \ffi (x) , \qquad \ffi' \equiv \ffi (x')  . 
\eeq
From  Eq.~\eqref{fifi} and \equref{deltaro} we have that the total change in the 
expectation of the two point operator is 
\beql{C1}\begin{split}
\tr ( \ffi \ffi' \delta\ro) & =  \tr  ( \ffi_L \ffi'_L \; \ro_{vac}^L \otimes \delta \ro^R)+
\tr (\ffi_R \ffi'_R  \; \ro_{vac}^L \otimes \delta \ro^R ) +
\tr ( \ffi_L \ffi'_R \delta\ro^e ) + \tr (\ffi_R \ffi'_L \delta \ro^e ) \\
& = \tr_L  ( \ffi_L \ffi'_L \ro_{vac}^L ) \tr_R \,  (\delta \ro^R)+
\tr_R \; (\ffi_R \ffi'_R \delta \ro^R ) + \tr ( \ffi_L \ffi'_R \delta\ro^e ) + \tr (\ffi_R \ffi'_L \delta \ro^e )   .
\end{split}\eeq
To get \equref{C1} we have used that $\tr (\hat{A}_{L,R} \; \ro^e) =0$ (see \equref{olrf}) and that 
\beql{tlfrl}
\tr_L ( \ffi_L \; \ro_{vac}^L) = \sum_J \, E_J^2 \, {}_L \! \bras{J}\ffi_L \kets{J}\! {}_L =0  ,
\eeq
since  states with different number of particles are orthogonal.

\equref{C1} is the farthest that we can get to reduce $\tr(\ffi\ffi' \delta \ro)$ using only
the fact that we are considering a source with support totally contained in $R$ and the properties of the matrix of entanglement. Note that the last two terms in
the \rhs{} of \equref{C1} are zero when evaluating both  $x,x' \in L$ or $R$ since in these wedges the modes $L_\omega$ and $R_\omega$ cannot be different from zero simultaneously. To go further in our calculation we shall introduce the explicit form of the operator $\hat{S}$,  which we do in the next section.

\section{An accelerated scalar source and its interaction}
\label{sec:QR}
We are going to use a scalar source $j(x)$ to model a \emph{scalar particle with uniform acceleration}. 
Let this scalar  current $j(x)$ interact with the field with an  interaction Hamiltonian density given by
\beql{hachei}
\hat{H}_I (x) =  \sqrt{-g} \,   j(x) \hat{\phi} (x)  ,
\eeq
where $g$ is the determinant of the metric.
The  final state $\kets{f}$ of the field \emph{after} 
this interaction is given by the application of the S-matrix to the inertial vaccuum state:
\beql{efin}
\kets{f}=\hat{S}\km \qquad \hat{S}=\hat{\mathrm{T}} \exp\left[ - i \int_{\Sigma_{in}}^{\Sigma_{out}} \hat{H}_I (x) \, d^2 x \right]  ,
\eeq
where $\Sigma_{in}$, $\Sigma_{out}$ are Cauchy  hypersurfaces where the interaction begins and ends respectively, and  $\hat{\mathrm{T}}$ is the time order operator.

The interaction occurs inside wedge $R$ and thus the \textit{in} and \textit{out}  Cauchy hypersurfaces should  bound this region and, at the same time, in order to define states in the Unruh quantization scheme, they have to be  
Cauchy hypersurfaces of the double wedge $L\cup R$.
We define  $\Sigma_{in}$ as the surface constructed by the union of $\{t=0,z\leq 0\}$ and
a spatial surface inside wedge $R$ which begins in the bifurcation point of the horizons and deviates slightly from the $\zeta=0$, $\tau=-\infty$ horizon (see
Fig.~\ref{fig1}). $\Sigma_{out}$ is defined analogously but its restriction to $R$ is 
an spatial surface which deviates slightly from the $\zeta=0$, $\tau=\infty$ horizon.
The initial vacuum state is defined over $\Sigma_{in}$ and the final state of the field  $\kets{f}$ is defined over $\Sigma_{out}$. 
Therefore, although the interaction is present inside wedge $R$, the state $\kets{f}$ is defined \emph{after} the interaction and one can evaluate expectations in this state of operators localized in wedges $R$ and $F$.

Our  calculation
of $\kets{f}$ (and its density matrix) will be in terms of Rindler coordinates for which 
$\hat{\mathrm{T}}$ orders up operators with respect 
to the time coordinate  $\tau$. 
Using \equref{hachei} one can put the 
final state of the interaction in the  form
\beql{f}
\kets{f}  =  \mathrm{T}\Big( \hat{1} -i  \int d^4 x \, \sqrt{-g} \,  j(x) \hat{\phi} (x) 
-\frac{1}{2} \int \! \!
\int  d^4 x \, d^4 x' \, \sqrt{-g}\sqrt{-g'} j(x') j(x') \hat{\phi} (x)\hat{\phi} (x')
\Big) \kets{0_M}  + \hat{O}(q^3)   ,
\eeq
where the integrations are over the same region as in \equref{efin}.
It is useful  to define the (formal) operator
\beq\label{fii}
\hat{\phi}_I \equiv  \int d^2x \, \sqrt{-g} \, j(x)\hat{\phi}(x)   .
\eeq
Note that  $\hat{\phi}_I$ is of order $q$.
We are going to apply the Wick theorem to the \rhs{} of Eq.~\eqref{f}:
\beql{wick}
\mathrm{T} \big( \hat{\phi}(x) \hat{\phi}(x') \big) = \mathrm{N} \big( \hat{\phi}(x) \hat{\phi}(x') \big) +
\bras{0} \mathrm{T}\Big(\hat{\phi}(x) \hat{\phi}(x') \Big)  \kets{0}  ,
\eeq
where the normal ordering, $\mathrm{N}$, and the vacuum, $\kets{0}$, are with respect to the  quantization scheme we are dealing with.
Using Eq.~\eqref{wick} we expand  Eq.~\eqref{f}  to get 
\beql{ho}
\kets{f} = \hat{S}\km=(1- \mathscr{G}) \km - i \hat{\phi}_I \km -\frac{1}{2} \mathrm{N} (\hat{\phi}_I  \hat{\phi}_I ) \km + \mathcal{O}(q^3)  ,
\eeq
where 
\beql{ggaril}
\mathscr{G} = \frac{1}{2}\int_{-\infty}^\infty d^2 x \int_{-\infty}^\infty d^2 y \; j(x) j(y) \bm \mathrm{T}\Big(\hat{\phi}(x) \hat{\phi}(y) \Big)  \km
\eeq
is of second order in $q$.

A uniformly accelerating scalar particle  follows a trajectory
which in Rindler coordinates corresponds to the locus of $\zeta= \zeta_0$.
From the point of view of Rindler observers,  this  scalar source  corresponds to a static scalar \emph{current}:
\beql{jacc}
j' (x) = q \delta (\zeta - \zeta_0)  .
\eeq
This source transforms as a scalar, using \equref{ccor} it can be transformed to  inertial 
coordinates  
\beql{jiner}
j' (x) = q  \frac{ \delta ( z - \sqrt{t^2 + a^{-2}})}{a\sqrt{t^2 + a^{-2}}} ,
\eeq
where $\zeta_0 = 1/a$ and $a$ is the proper acceleration of the source.
In order to avoid expressions of the form $0\times\infty$  one has to introduce some regularization factor to  Eq.~\eqref{jacc}. 
Inspired in 
Higuchi \etal{} \cite{bremstralung}, who regularize an electric charge and show the consistency of their regulator 
(we have explained their results in Sec.~\ref{sec:intro}), 
we introduce an oscillating factor to Eq.~\eqref{jacc}:
\beql{jaccr}
j(x) = q  \cos (\th \tau)  \delta (\zeta - \zeta_0)
\eeq
and at the end of our calculations we shall take the limit $\th\to 0$.
In fact,  for slow oscillations ($\th\ll a$) the source is expected to interact with the field as if it were a constant charge $q$ at each $\tau$.

The current $j(x)$ has support totally contained in $R$, and thus 
one can see  from Eqs.~\eqref{fiun} and~\eqref{fii}  that  only $R$-modes of the field will get excited by the accelerated source.
The operator $\hat{\phi}_I$ takes  the form
\beql{fiii}
\hat{\phi}_I =  \int_R d^2x \, \sqrt{-g} \, j(x)\hat{\phi}_R (x)=
 \int_0^\infty d\omega \, \Big( \Upsilon_\omega \hat{r}_\omega + \Upsilon_\omega^* \hat{r}_\omega^\dagger \Big)  ,
\eeq
where
\beql{psith}
\Upsilon_\omega \equiv\int_R d^2 x \, \sqrt{-g} j(x) R_\omega (x)=
\Psi_\th \delta (\omega -\theta )  ,
\qquad 
\Psi_\th = q\sqrt{\sinh(\pi\theta)}\, \zeta_0 K_{i\theta} (m\zeta_0)  .
\eeq
To get \equref{psith}  we have used 
Eq.~\eqref{errew},  $\omega>0$,  and we have chosen to work with   $\theta>0$ (note that the  regulator is an even function of $\th$).
Note that   $K_{i\omega} (z)$ is real for 
real $\omega$ and $z$. Finally, Eq. \eqref{fiii} takes  the form
\beql{fii2}
\hat{\phi}_I = q\sqrt{\sinh(\pi\theta)}
\,  \zeta_0 K_{i\theta} (m\zeta_0) \big[ \hat{r}_\theta + \hat{r}_\theta^\dagger \big]  .
\eeq
The role of the regulator we have chosen is to couple  the 
source to an  $R$ mode of the field with frequency $\th$ instead of 
the mode with frequency zero which is somehow pathological.

\section{The change of $\ev{\hat{T}_{\mu\nu}}$ at wedges $L$ and $R$}
\label{sec:eval}

Now we are in a position to compute  explicitly
$\delta\ro^R$ and $\delta\ro^e$  in order to evaluate 
\equref{C1}. 
We will work out this calculation perturbatively up to second order in $q$, which turns to be the first relevant contribution.
From \equref{ho} we have 
\beql{essse}
\hat{S} = (1- \mathscr{G}) -i\ffi_I +\frac{1}{2} N(\ffi_I \ffi_I ) + O(q^3 )  ,
\eeq
and from this equation we can write the density matrix for the final state as
\beql{rof012}
\ro_f = \ro_f^{(0)}+\ro_f^{(1)}+\ro_f^{(2)}+O(q^3)  ,
\eeq
where $\ro_f^{(0)}$ corresponds to the (second order) renormalized inertial vacuum density matrix given by
\beql{ro0}
\ro_f^{(0)} = \mathit{Q} \ro_{vac} \qquad
\mathit{Q}\equiv (1-2\mathrm{Re}(\mathscr{G})) .
\eeq
We shall compute the change in the density matrix 
with respect to the renormalized vacuum density operator:
\beq
\delta \ro_{ren} = \ro_f - \ro_f^{(0)}  .
\eeq
Tracing out the left degrees of freedom in \equref{rof012} we have
\beq
\ro_f^R =  \ro_f^{R\,(0)}+ \ro_f^{R\,(1)} + \ro_f^{R\,(2)} + O(q^3)  ,
\eeq
where 
\beqar
\ro_f^{R\,(1)} & = & -i [\ffi_I , \ro_{vac}^R ] , \label{rfr1}\\
\ro_f^{R\,(2)} & = & \frac{1}{2} \big( \ro_{vac}^R N(\ffi_I \ffi_I )^\dagger 
+ 2 \ffi_I    \ro_{vac}^R \ffi_I + N(\ffi_I \ffi_I ) \ro_{vac}^R \big)\label{rfr2}  . 
\eeqar
Note that $\ro_f^{R\,(1)}$ is traceless. For the entanglement matrix  we have
\beq
\delta\ro^e = \ro_f^{e\,(1)} + \ro_f^{e\,(2)} + O(q^3) ,
\eeq
where $\ro_f^{e\,(1)}$ and $\ro_f^{e\,(2)}$ are defined as Eqs.~\eqref{rfr1} and \eqref{rfr2} changing $R\to e$.

The first order contribution to \equref{C1} is
\beql{prim}
\tr (\ffi \ffi' \delta\ro^{(1)})=  \tr_L (\ffi_L \ffi'_L \ro_{vac}^L)\tr ( \ro_f^{R\,(1)}) +
\tr (\ffi_R \ffi'_R  \ro_f^{R\,(1)} ) + \tr ( (\ffi_L\ffi'_R + \ffi'_L \ffi_R ) \ro_f^{e\,(1)})  .
\eeq
The first term in the \rhs{} of \equref{prim} is zero since $\ro_f^{R\,(1)}$ is traceless.
From the expression for $\ro_f^{R\,(1)}$,  \equref{rfr1}, it can  be proved that
\beql{st}
\tr(\ffi_R \ffi'_R \ro_f^{(1)}) = \tr(\ffi_L \ffi'_R \ro_f^{(1)}) = \tr(\ffi_R \ffi'_L \ro_f^{(1)}) = 0  .
\eeq
This can be understood heuristically from the fact that   these traces
represent a sum of brackets in the Minkowski vacuum of three field operators, which are necessarily null. From \equref{st}  and the definition of $\ro^e_f$ we have 
\beql{tffdre2}\begin{split}
\tr (\ffi_L \ffi'_R  \ro_f^{e\,(1)}) & = \tr(\ffi_L \ffi'_R \ro_f^{(1)}) 
- \tr ( \ffi_L \ffi'_R \ro_{vac}^L \otimes \ro_f^{R\, (1)}) \\
& = \tr(\ffi_L \ffi'_R \ro_f^{(1)}) \\
& = 0  .
\end{split}
\eeq
To get the second equality in \equref{tffdre2} we have used  \equref{tlfrl} to conclude that
\beql{511}
 \tr ( \ffi_L \ffi'_R \; \ro_{vac}^L \otimes \ro_f^{R\, (1)})
 = \tr_L \, (\ffi_L \ro_{vac}^L) \tr_R \, (\ffi'_R \ro_f^{R\, (1)}) = 0  \,.
 \eeq
Note that this last equation is valid for any order.
We have then proved that $\tr ( \ffi \ffi' \delta \ro^{(1)}) = 0 $.

Now we turn to the second order contribution,
\beql{segu}
\tr (\ffi \ffi' \delta\ro^{(2)})=  \tr_L (\ffi_L \ffi'_L \ro_{vac}^L)\tr ( \ro_f^{R\,(2)}) +
\tr (\ffi_R \ffi'_R  \ro_f^{R\,(2)} ) + \tr ( (\ffi_L\ffi'_R + \ffi'_L \ffi_R ) \ro_f^{e\,(2)})  \,.
\eeq
For the mixed $LR$ terms,  we have that, similarly to \equref{tffdre2} 
\beql{LR2}\begin{split}
\tr (\ffi_L \ffi'_R \; \ro_f^{e\,(2)}) & = \tr(\ffi_L \ffi'_R \; \ro_f^{(2)}) 
- \tr ( \ffi_L \ffi'_R \; \ro_{vac}^L \otimes \ro_f^{R\, (2)}) \\
& = \tr(\ffi_L \ffi'_R \; \ro_f^{(2)})  \,. \\
\end{split}
\eeq
Using \equref{rfr2} it can be proven directly that 
(see Appendix~\ref{app:scnd})
\beqar
\tr (\ffi_L \ffi'_R  \ro_f^{e\,(2)})  & = & Z\Psi_\th^2 \tr(\ffi_L \ffi'_R \ro_{vac} )\label{tffdre3},\\
 \tr(\ffi_ R \ffi'_L \ro_f^{e\,(2)})  & = & Z\Psi_\th^2 \tr(\ffi_R \ffi'_L \ro_{vac} )\label{tffdre4}\,.
\eeqar
Also from \equref{rfr2} (see its basis expansion in \equref{ror2ap}) we have that
$\tr(\ro_f^{R\,(2)})=\Psi_\th^2$ and thus 
\beql{pffl}
\tr_L (\ffi_L \ffi'_L \ro_{vac}^L)\tr ( \ro_f^{R\,(2)}) = Z \Psi_\th^2 \tr_L (\ffi_L \ffi'_L \ro_{vac}^L)\,.
\eeq
These terms, since are proportional to the corresponding $LR$ and $LL$ parts of the two point function in the vacuum, are going to be absorbed in the renormalization of the final result.
We have  then obtained   the, in principle, unexpected  result that at least for the change in the 
expectation of the $\hat{T}_{\mu\nu}$ operator \emph{there is no contribution 
from the change in the entanglement matrix}, although as one can see from
\equref{roefe} there is actually a change in this operator.  
All the information of this change  comes from the change in the 
$R$ density matrix of the state, $\delta\ro_f^R$.
As shown explicitly in Appendix~\ref{app:scnd} we have that
\beql{ro2rr}
\tr( \ffi_R \ffi'_R  \ro_f^{R(2)} ) = 4 \Psi_\th^2 \,  \mathrm{Im}[R_\th (x)]\: \mathrm{Im}[R_\th (x')] + 
Z \Psi_\th^2 \tr \big( \ro^R_{vac} \ffi_R (x) \ffi_R (x')  \big) \,.
\eeq
Finally, adding up Eqs.~\eqref{tffdre3}-\eqref{ro2rr}
we have that the change in the two point function between the inertial vacuum and the state generated by the interaction of the scalar source is 
\beql{C2}
\tr ( \ffi(x) \ffi(x') \delta \ro_{ren} ) =  4 \Psi_\th^2 \,   \mathrm{Im}[R_\th (x)]\: \mathrm{Im}[R_\th (x')] + 
\frac{Z \Psi_\th^2}{\mathit{Q}} \tr \big(  \ffi (x) \ffi (x') \ro^{(0)}_{f}  \big)+ O(q^3) ,
\eeq
which is valid for all $x,x' \in M$ and we have used \equref{ro0} to express
$\ro_{vac}$ in terms of the renormalized vacuum  density operator $\ro_f^{(0)}$. However, as we have said, we still have to renormalize
this change in the expectation value. We define  
 the renormalized field operator as 
\beq
\ffi_{ren} (x) \equiv \left( 1- \frac{\Psi_\th^2}{\mathit{Q}}\right)^{1/2} \ffi(x),
\eeq
and thus  the renormalized change  in the expectation of the two point function is 
\beql{DD}
C_\th (x,x') \equiv \tr ( \ffi_{ren} (x) \ffi_{ren} (x') \delta \ro_{ren} ) = 4 \Psi_\th^2 \,   \mathrm{Im}[R_\th (x)]\: \mathrm{Im}[R_\th (x')] + O(q^3 ) \,.
\eeq
The second order term in this change reads
\beql{DD2}
C_\th^{(2)} (x,x') = 4 \Psi^2_\th \,\mathrm{Im}[R_\th (x)]\: \mathrm{Im}[R_\th (x')]\,.
\eeq

We have arrived to a regular expression for the change in the expectation value of the two point function and thus now we can reconsider the source as static by taking out the regulator taking the limit $\th\to 0$.
In the following we are going to evaluate explicitly \equref{DD2} in wedges $R$ and $F$. 
First, note that form \equref{psith} and the fact that 
Bessel functions $K_{i\th}(m\zeta_0)$ are regular whenever $\zeta_0 \neq 0$ we have that
\beql{psi2lim}
\lim_{\th\to 0} \Psi_\th = 0  \,.
\eeq
Note that $\Psi_\th^2$  is an overall factor in $C_{\th}^{(2)}(x,x')$.  
When computing the expectation of $\ffi(x)\ffi(x')$ the correct procedure is to take the limit $\th\to 0$ at 
the end of the calculation since the field operators \emph{couple} to the frequency $\th$ of particles in $\kets{f}$ 
which are emitted and  absorbed by the source. 

First we will work the case when  $x,x'\,\in R$, for we can express these points in
terms of Rindler coordinates: $x=(\tau,\zeta)$, $x'=(\tau',\zeta')$.
Unruh modes $R_\omega$, when restricted to wedge $R$ take the form of Eq.~\eqref{fullmod} (see also Eq.~\eqref{errew} in the Appendix).  
Using the fact that  functions $K_{i\omega}(z)$ are real whenever $\omega$ is real and $z>0$, we have that
\beq
\mathrm{Im} \big( R_{\th} (x) \big) = -\frac{\sqrt{\sinh (\pi\th)}}{\pi} \sin (\th\tau) K_{i\th}(m\zeta) 
\qquad x\in R \,.
\eeq
And thus, from Eqs.~\eqref{DD2} and~\eqref{psith}
\beq
C_\th^{(2)} (x,x') = \frac{1}{\pi^2} q^2 \sinh^2 (\pi\th) \zeta_0^2 K_{i\th}^2 (m\zeta_0) \sin (\th\tau)\sin (\th\tau')
K_{i\th} (m\zeta) K_{i\th}(m\zeta') \qquad x,x'\in R \,.
\eeq
Recall that $R$ is an open set bounded by the horizons and thus $\zeta,\zeta' \neq 0$, so the Bessel 
functions $K_{i\th} (m\zeta)$,  $K_{i\th}(m\zeta')$ are regular. Then we have that the second order change
in the two point function between the inertial vacuum and state $\kets{f}$ in wedge $R$ is 
\beql{DR}
\lim_{\th\to 0} C_\th^{(2)} (x,x') = 0 \qquad x,x'\in R  \,.
\eeq
This resut, is consistent with the fact that, as pointed out in Ref.~\cite{bremstralung} the source is 
in thermal equilibrium with the field inside wedge $R$. 
Any observer inside this wedge will not be able to
notice any change in the expected value of $\hat{T}_{\mu\nu}$ due to  the presence of the accelerating source. Note that the specific form of the two point operator has played a crucial role to get to \equref{DD2}.

Now we proceed to evaluate Eq.~\eqref{DD2} in wedge $F$. Recall that in $F$, $\tau$ is a spatial coordinate
and $\zeta$ is timelike (\cf{} Eq.~\eqref{cF}).
Form  Eq.~\eqref{erreF} we have that  the mode $R_\th (x)$ restricted to $F$ takes the form
\beql{RF}
R_\th  (x) = - \frac{i}{2^{3/2}} \frac{e^{-i\th\tau}}{\sqrt{2\sinh(\pi\th)}} 
\left[ e^{\th\pi}H_{i\th}^{(2)} (m\zeta) + e^{-\th\pi} H_{i\th}^{(1)} (m\zeta) \right]
\qquad x\in F,
\eeq
where coordinates $(\tau,\zeta)$ are defined in \equref{cF}  and  $H_{i\theta}^{(1),(2)}$ are the first and
second  Hankel functions. 
Using the definitions of $H_{i\theta}^{(1),(2)}$ in terms of Bessel functions,
Eqs. \eqref{h1} and \eqref{h2}, it  follows immediately that
\beql{ImRF}
\mathrm{Im}\big( R_\th  (x) \big) = - \frac{1}{4\sqrt{\sinh(\pi\th)}}
\big( e^{-i\th\tau} J_{-i\th} (m\zeta) + e^{i\th\tau}J_{i\th} (m\zeta) \big) \qquad x\in F \,.
\eeq
Now using Eqs.\eqref{psith} and \eqref{ImRF} we have that
\beql{DF}
\lim_{\th\to 0} C_\th^{(2)} (x,x') =  q^2 \zeta_0^2 K_{0}^2 (m\zeta_0) J_0 (m\zeta) J_0 (m\zeta') 
\qquad \text{for all } x,x' \in F \,.
\eeq
Compare with \equref{DR}. This expression is the (non zero) change in the two point function in wedge $F$ 
due to the interaction. It contains the information of the field radiated away from the
source into $F$. It should be noted that \equref{DF} is not valid at the horizons.
In Appendix~\ref{app:inertial} we make the computation, in an inertial frame, of  the same change in the expectation of the two point function, $\tr(\ffi(x)\ffi(x')\delta\ro)$, for $x,x'\in F$ and obtain exactly \equref{DF}. Therefore, at least for the particular case we are dealing with, both quantum descriptions produce the same physical results.

\section[Left-right interference]{An example of left-right interference: two sources with opposite acceleration}
\label{sec:ex}
As we have seen in the last section, the matrix of entanglement plays no significant role in the change in the expectation of $\hat{T}_{\mu\nu}$ for the interaction of an accelerated source with the inertial vacuum. However, the presence of another (accelerating) source lying in $L$ may affect the entanglement of the final state and thus $\ro^e$ may play a part in the change of the expectation of $\hat{T}_{\mu\nu}$.
In this section we work out these calculations.

Suppose  that additionally to the scalar particle considered in the latter sections we have an extra scalar particle in uniform acceleration in wedge $L$ with scalar charge $\tilde{q}$.
The  perturbation of the field due to this pair of particles  is given by
\beql{jlr}
j(x) = \left\{
\begin{array}{ll}
 j_L (x)  & x\in L \\
 j_R (x)  & x\in R
 \end{array}\right. ,
\eeq
where
\beq
j_L (x) =  \tilde{q}\cos ( \tilde{\theta} \, \tau_L ) \delta (\zeta_L - \tilde{\zeta}_0 ) \qquad 
j_R (x) = q\cos (\theta \,\tau_R ) \delta (\zeta_R - \zeta_0 ) \,.
\eeq
Rindler coordinates in wedge $L$, $(\tau_L,\zeta_L)$ are given by \equref{cL}. To avoid confusion, all along this section we will
call Rindler coordinates in $R$: $(\tau_R,\zeta_R)$ (originally we have used  $(\tau,\zeta)$).
Note that in order to have independent sources we have introduced the cosine regulator with a different parameter for the source in wedge $L$.

Using \equref{hachei} for $j(x)$ given by \equref{jlr} we have that
\beql{slr1}
\hat{S}' = \hat{T} \exp \big[ -i [(\ffi_I^L \otimes \hat{1}_R)+ (\hat{1}_L \otimes \ffi_I^R )]
\big]
= \hat{T} \;  \exp [ -i \ffi_I^L ] \otimes \exp [ -i \ffi_I^R ]
 \equiv  \hat{S}_L \otimes \hat{S}_R ,
\eeq
where
\beq
\ffi_I^L = \Psi_{\tilde{\th}} \big( \hat{l}_{\tilde{\th}} + \hat{l}_{\tilde{\th}}^\dagger \big),
\qquad
\ffi_I^R = \Psi_\th \big( \hat{r}_\th + \hat{r}_\th^\dagger \big),
\eeq
and $\hat{T}$ is the time order operator. In this case we have 
two different time parameters, $\tau_L$ and $\tau_R$  and thus $\hat{T}$ will time order
$L$ and $R$ operators independently (recall that $\ffi_L$ and $\ffi_R$ commute).
The factor $\Psi_\th$ is given by  \equref{psith} and $\Psi_{\tilde{\th}}$ is now
\beq
\Psi_{\tilde{\th}} = \tilde{q}\sqrt{\sinh(\pi\tilde{\theta})}\, \tilde{\zeta}_0 K_{i\tilde{\theta}} (m\tilde{\zeta}_0)\,.
\eeq

Let $\kets{g}$ be the final state of this interaction, analogously to \equref{SS} now we have that its density matrix takes the form
\beq
\ro_g = \ro_g^L \otimes \ro_g^R + \ro_g^e ,
\eeq
where
\beq
\ro_g^L = \hat{S}_L \ro_{vac}^L \hat{S}_L^\dagger ,
\qquad \ro_g^R = \hat{S}_R \ro_{vac}^R \hat{S}_R^\dagger , 
\qquad
\ro_g^e = \hat{S}' \ro_{vac}^e \hat{S}^{\prime\dagger} \,.
\eeq
Then we have that  the change in the density operators is given by
\beql{dlr}
\delta \ro_g 
 = \ro_{vac}^L \otimes \delta \ro_g^R +
\delta \ro_{g}^L \otimes  \ro_{vac}^R + \delta \ro_{g}^L \otimes \delta \ro_{g}^R
+ \delta \ro^e_g  ,
\eeq
where  $\delta\ro_g = \ro_g - \ro_{vac}$ and all other differences  are defined analogously.

Now we compute the change $\tr ( \ffi \ffi' \delta \ro_g )$, analogously to the case of the single accelerating particle the first relevant term is of second order in $q$ and $\tilde{q}$.   Using \equref{fifi} and  \equref{dlr} It can be seen that
\beql{ffdlr}\begin{split}
\tr (\ffi \ffi' \delta \ro_g^{(2)} ) = & 
\tr_L \, ( \ffi_L \ffi'_L  \delta \ro_g^{L\,(2)} ) + \tr_R \,  ( \ffi_R \ffi'_R  \delta \ro_g^{R\,(2)} )\\
{} & + Z\Psi_\th^2 \tr_L \, ( \ffi_L \ffi'_L \ro_{vac}^L ) 
	+Z\Psi_{\tilde{\th}}^2 \tr_R \, ( \ffi_R \ffi'_R \ro_{vac}^R ) \\
{} & + \tr\big( ( \ffi_L \ffi'_R + \ffi_R \ffi'_L ) \: \delta\ro_g^{L\,(1)}\otimes\delta\ro_g^{R\,(1)} \big)\\
{} & + \tr \big( ( \ffi_L \ffi'_R + \ffi_R \ffi'_L ) \delta \ro_g^e \big) \,.
\end{split}
\eeq
The first four terms in the \rhs{} of \equref{ffdlr} are analogous to \equref{pffl} and 
and \equref{ro2rr} for the single source case. In effect, we have that
\beql{ro2Rdc}
\tr( \ffi_R \ffi'_R  \delta\ro_g^{R(2)} ) = 4 \Psi_\th^2 \,  \mathrm{Im}[R_\th (x)]\: \mathrm{Im}[R_\th (x')] + 
Z \Psi_\th^2 \tr \big( \ro^R_{vac} \ffi_R (x) \ffi_R (x')  \big) ,
\eeq
\beql{ro2Ldc}
\tr( \ffi_L \ffi'_L  \delta\ro_g^{L(2)} ) = 4 \Psi_{\tilde{\th}}^2 \,  \mathrm{Im}[L_{\tilde{\th}} (x)]\: \mathrm{Im}[L_{\tilde{\th}} (x')] + 
Z \Psi_{\tilde{\th}}^2 \tr \big( \ro^L_{vac} \ffi_L (x) \ffi_L (x')  \big) \,.
\eeq
Similarly to \equref{rfr1}, it can be seen that 
\beqar
\delta\ro_g^{L\,(1)} & =&  -i ( \ffi_I^L \ro_{vac}^L - \ro_{vac}^L \ffi_I^L ) ,
\label{drgl1}\\
\delta\ro_g^{R\,(1)} &=&  -i ( \ffi_I^R \ro_{vac}^R - \ro_{vac}^R \ffi_I^R )  ,
\label{drgr1}
\eeqar
and from these equations  we have that  
\beql{inter}
 \tr\big( ( \ffi_L \ffi'_R + \ffi_R \ffi'_L ) \: \delta\ro_g^{L\,(1)}\otimes\delta\ro_g^{R\,(1)} \big)
 =4 \Psi_{\tilde{\th}} \Psi_\th \big( \mathrm{Im}
(L_{\tilde{\th}} (x) ) \mathrm{Im} ( R_\th (x') ) + \mathrm{Im}(R_{\tilde{\th}} (x) ) \mathrm{Im} ( L_\th (x') ) \big) \,.
 \eeq
On the other hand, it can be shown that the second order contribution from the matrix of entanglement is
\beq
\tr \big( ( \ffi_L \ffi'_R + \ffi_R \ffi'_L ) \delta \ro_g^{e\,(2)} \big)= Z
\big( \Psi_{\tilde{\th}}^2 + \Psi_{\th}^2 \big) \;
\tr \big( ( \ffi_L \ffi'_R + \ffi_R \ffi'_L ) \ro_{vac} \big)  ,
\eeq
which, as in the case of a single source, also corresponds to terms that will be 
absorbed in the renormalization of the final change in the expectation value.
Adding up all the contributions, \equref{ffdlr} takes the form
\beql{ffdrl2}\begin{split}
\tr (\ffi \ffi' \delta \ro_g^{(2)} ) = &
4\big[ \Psi_{\tilde{\th}} \mathrm{Im}( L_{\tilde{\th}} (x)) + \Psi_{\th} \mathrm{Im}
(R_\th (x) )
\big]\big[ \Psi_{\tilde{\th}} \mathrm{Im}( L_{\tilde{\th}} (x')) + \Psi_{\th} \mathrm{Im}
(R_\th (x') )
\big]  \\
& + Z\big( \Psi_{\tilde{\th}}^2 + \Psi_{\th}^2 \big) \;
\tr \big(  \ffi (x) \ffi (x') \ro_{vac} \big)  \,.
\end{split}
\eeq
Thus, the renormalized change in the two point function reads, up to second order
\beq
\tr (\ffi_{ren} \ffi'_{ren} \delta \ro_{g\, ren} )  =
4\big[ \Psi_{\tilde{\th}} \mathrm{Im}( L_{\tilde{\th}} (x)) + \Psi_{\th} \mathrm{Im}
(R_\th (x) )
\big]\big[ \Psi_{\tilde{\th}} \mathrm{Im}( L_{\tilde{\th}} (x')) + \Psi_{\th} \mathrm{Im}
(R_\th (x') )
\big]  + \dots 
\eeq
 After taking the limits $\tilde{\th},\th \to 0$ and evaluating
at $x=(\tau_F , \zeta_F )$ and $x'=(\tau'_F , \zeta'_F )$ we obtain
\beql{ffrgfin}
\lim_{\tilde{\th},\th\to 0} \tr (\ffi_{ren} (x)  \ffi_{ren} (x') \delta \ro_{g\, ren} )  =
 \big( \tilde{q}\tilde{\zeta}_0 K_0 (m\tilde{\zeta}_0 ) + q\zeta_0 K_0 (m\zeta_0 ) \big)^2
J_0 (m\zeta_F ) J_0 (m\zeta'_F ) \,.
\eeq
As expected, if we turn off the charge in wedge $L$ ($\tilde{q}\to0$) we recover our previos result, \equref{DF}. For the case of the two accelerating sources it turns out that also all the contribution to the change in the expectation value of the two point function comes solely from the change in the partial matrices, $\delta\ro_g^L$, $\delta\ro_g^R$.
In particular, the interference term, \equref{inter}, is determined by the latter pair of matrices, that is, for the case we have just analyzed all the information of the change in the expectation of $\hat{T}_{\mu\nu}$ is \emph{only} encoded in 
$\delta\ro_g^L$, $\delta\ro_g^R$.

\section{Discussion}
\label{sec:dis}

One of the main goals of our work was to reconcile the fact that  the final state of the field appears to remain  an undisrupted  thermal state in both the left and right Rindler wedges, with the expected change  induced by the source on   field observables, such as the energy momentum, in the  future wedge.
This issue is recasted in terms of  the  codification in the  state
of the field of  the  pertinent information  that exhibits the  change in the expectation value of $\hat{T}_{\mu\nu}$ .

At the beginning it was our belief that, since there is no change in the expectation of $\hat{T}_{\mu\nu}$ in wedge $R$ (neither in $L$) the physical change in wedge $F$ cannot be induced by the particular behaviours of the state when restricted to either wedge. Hence, the information of this change  information should have been encoded in some part of the state which is not represented by any of the restricted density operators $\ro_f^L$ and $\ro_f^R$ (or in their respective changes). It is in this sense is that we proposed the decomposition of $\ro_f$ given by \equref{ro+roe} with the particular introduction of the \emph{matrix of entanglement} $\ro^e$. As we explained in Sec.~\ref{sec:pos} this operator plays no role when computing expectations of observables localized in wedges $L$ and $R$ and thus was a good candidate to account for the change of $\ev{\hat{T}_{\mu\nu}}$ in wedge $F$.
Nevertheless, we computed  this change perturbatively and found that
it has contributions only from  the change  in the density matrix describing the state in the wedge $R$ (see Eqs.~\eqref{ro2rr} and~\eqref{DD}).
That is,  when  evaluating  $\ev{\hat{T}_{\mu\nu}}$ in either wedges $R$ and $F$, its change is determined solely by the characterization  of the state  in wedge $R$. 
This result contrasts with  our initial expectation that the information about 
the change of $\ev{\hat{T}_{\mu\nu}}$ would  be encoded in the change in the entanglement matrix, $\delta\ro^e$. 

In order to obtain this result, we had to introduce by hand a particular regularizing function, $\cos (\th \tau )$,  into the current describing a uniformly accelerating scalar source with the prescription to take the limit $\th\to 0$ at the end of the calculation.  As mentioned earlier, this regulator was inspired by a similar computation done in~\cite{bremstralung}.
Despite the seemingly artificial choice of this regulator, we have shown 
that when  making the regulator independent calculation of  the change in the same expectation $\ev{\ffi(x)\ffi(x')}$, using a plane wave quantization scheme,  one obtains the same results as in the Unruh scheme with such  regulator (see Appendix~\ref{app:inertial}).
One can give an heuristic explanation to the fact that 
 this regulator is physically correct as follows.
In principle, if one is  precise, one would like to describe the radiation due to a real physical particle,
which should be described  as a quantum object itself.  However, this
description has a serious drawback regarding our wish to describe the source as a \emph{uniformly accelerating} particle: A quantum particle does
not move in  a definite trajectory and thus assigning to it a particular
acceleration is impossible. On the other hand, the nature of the quantum field is
distributional and therefore the correct quantum description of the interacting
source should be in terms of test functions of compact support (note that
\equref{jaccr} lacks this property) and therefore, the source would
correspond to an extended object.  To such object we can not  naturally ascribe a \emph{uniform} acceleration: if the object is to maintain ``its shape''
along its trajectory then different parts should have different proper
accelerations. 
However, we know from  \cite{bremstralung} that  a treatment using classical point-like sources (with  definite
proper acceleration)   together with a certain type of regulator produce physically correct results (in particular,
 results that are fully consistent with the Equivalence Principle).

The regulator used in \cite{bremstralung} consisted in the introduction of an artificial oscillation  with frequency $\theta$ in the strength of the source, to identify therefore expressions of the form  $0\times \infty$ occurring in the calculation and  to proceed to carry all calculations to the end before taking the limit $\theta \to 0$. 
We are thus  assuming  that the introduction of such regulator,
along with the prescription to take the  limit $\th\to 0$ at the end of  the calculations reflects  in an effective way the
description of a quantum source in uniform acceleration interacting with the
field.
(Nevertheless, the robustness of the result would be ensured
if one confirms that the same physical behaviour is to be obtained for a wider
class of regularizing functions.)
It is however worthy to emphasize  that in the inertial calculation of Appendix~\ref{app:inertial} there was no need to introduce such a regulator. 

From these considerations it follows that  our calculations make sense only if the limit $\th\to 0$ is
taken at the 
end of the computation of the expectation values. Actually, one may be tempted
to take this limit directly into the density matrix $\ro_f$ of the state.
Ignoring for a moment details regarding the precise notion of  limit of an operator, one can
see that due to  the overall factor 
$\Psi_\th$, which goes to zero as $\th\to 0$, every single term in the expansion
of $\ro_f$ is also zero except for those terms proportional to  $\ro_{vac}$. Thus, one would conclude that there has been no change in the state of the field in $R$ due to the presence of the source. As this was the only potential contribution to the  change in the expectation
of $\hat{T}_{\mu\nu}$, we would  be led to the erroneous conclusion that there is no change in this quantity.

From the calculations in Sect.~\ref{sec:eval} we concluded that the information of the change in $\ev{\hat{T}_{\mu\nu}}$ is encoded in $\delta \ro^R$. We  now want to consider   \emph{how}
is it encoded.  
The answer to this question relies on a subtle  interplay of the field operator
$\ffi(x)$
and  the density matrix  in the present formalism.
Let us focus on  the details  of our specific calculation:
The overall factor $\Psi_\th^2$ in the second order contribution to $\ro^R_f$
comes from the fact that the source is located in wedge $R$. In this wedge
Unruh modes $R_\th (x) \to 0$ as $\th\to 0$ (whenever $x$ is not at the horizon).
When computing $\tr (\ffi_R \ffi'_R \ro_f^{R\,(2)})$ the field operators are
sensitive to the frequency $\th$. 
In fact, they only excite modes with $\omega=\th$ as could be expected on Rindler energy conservation grounds.
The particular form of $\ro_f^R$ determines the
structure of the contribution given by \equref{DD2}  which in turn, due to the
different behaviors of the Unruh modes in wedges $R$ and $F$, is zero 
in the former and not zero in the latter.

We interpret these results  by saying that the ``0'' Rindler energy modes,
which were of such concern in regard to the definition of the theory (see discussion at the end of Sec.~\ref{subsec:bmum}), are in effect, essential in order to obtain in the accelerated frame description identical results as in the inertial one.  Physically we could think that these modes are excited by the  slightest quantum fluctuations of a realistic quantum particle and that their  excitation would be directly felt in the future wedge. These results seem to be in accordance with the spirit of those obtained in~\cite{bremstralung2} where it is argued that the zero energy modes seemed to be undetectable (with an appropriate definition of detectability) when confining the detection to the right wedge. 
Finally, it is our belief that this work has helped in clarifying the questions raised at the beginning. Furthermore, there is a technically analogous situation which
indicates that a stationary  particle just  outside of the horizon of a stationary black hole could be 
``emitting'' towards its interior. This work shows a clear path to studying the changes in the energy momentum tensor in that situation.

\begin{acknowledgments}
The authors would like to thank very helpful discussions with George Matsas and Hanno Sahlmann. 
This work was in part supported by grants
DGAPA-UNAM IN108103-3 and CONACYT 43914-F and 36581-E. I.P.~would like to thank the CONACYT Graduate scholarship and SNI-assistant grant. 
\end{acknowledgments}

\appendix
\section{Unruh modes and representations of boost modes}
\label{sec:repre}

One can generalize Rindler coordinates to all Rindler wedges 
assigning always  the respective coordinate $\tau$ to the 
parameter associated to the generator of  boosts about the origin in the $z$ direction:   
\beql{ba}
b^\mu = 
 a \left[ z \left(\frac{\partial}{\partial t}\right)^\mu + t \left(\frac{\partial}{\partial z}\right)^\mu \right] \,.
\eeq
In wedge $L$ one should do an exception in order to  to guarantee that the future direction coincides 
with that of inertial time, in this case,  $\tau^\mu$ is  $-b^\mu$. In wedges $F$ and $P$ the coordinate $\tau$ is spacelike. We have ~\cite{boulware_qntzn}:
\begin{subequations}\label{tcors}\begin{align}
z=&-\zeta \cosh (a\tau ) &   t&=\zeta\sinh (a\tau )   &(t,z) \in L &\label{cL}  ,\\
z=&\;\zeta\sinh (a\tau ) &   t&=\zeta \cosh(a\tau )   &(t,z) \in F &\label{cF}   ,\\
z=&-\zeta \sinh (a\tau ) &   t&=-\zeta\cosh (a\tau )  &(t,z) \in P &\label{cP},
\end{align}\end{subequations}
where  $\zeta>0$ in each region. 

Boost modes, \equref{bm},  should be thought of as distributions and thus one cannot evaluate them in one particular 
point. However, what we can do is to apply this distributions to test  functions which have support defined on a certain 
open region. The integral which defines the boost modes can be  expressed in  terms of the accelerated
 coordinates $(\tau,\zeta)$ given by Eqs. \eqref{ccor} and \eqref{tcors}. As Unruh modes are defined by boost modes, 
from this operation we can express Unruh modes in accelerated coordinates too.

For example,  let $f$ be a test function in $M$ with $\mathrm{supp}( f ) \subset R$, the evaluation
of $B_\omega$ at function $f$ is given by
\beq
B_\omega  [f]=\frac{1}{2^{3/2}\pi}\int_{-\infty}^\infty d\theta \; e^{-i\omega\theta} \int_R d^2 x\;   
e^{im \, (z\sinh(\theta)-t\cosh(\theta))} f(t,z)  \,.
\label{bmr}
\eeq
Since the space-time integration is in wedge $R$ we can express the $d^2 x$ integral  in terms of Rindler
coordinates (\mbox{c.f.} Eq.\eqref{ccor}):
\beq
B_\omega  [f]=\frac{1}{2^{3/2}\pi}\int_{-\infty}^\infty d\theta \; e^{-i\omega\theta} \int_R d^2 x\;   e^{im\zeta\sinh(\theta-\tau)}\, f(\tau,\zeta) \,.
\eeq
Changing the integration order and using the following relation for the Bessel functions~\cite{erdelyi} 
\beq
K_\nu (x)=\frac{1}{2} e^{\frac{i}{2}\nu\pi} \int_{-\infty}^{\infty}    d\alpha \  e^{\nu\alpha} e^{-ix\sinh(\alpha)} ,
\eeq
it can be seen that 
\beq
B_\omega [f] = \frac{1}{\pi\sqrt{2}} e^{\frac{\omega\pi}{2}} \int_R d^{\,2} x \, e^{-i\omega\tau} K_{i\omega}(m\zeta) f(\tau,\zeta) \,.
\eeq  
By analogous arguments it can be seen that in other wedges boost modes take the following
form \cite{gerlach} (coordinates in the following eqs. are respective to 
each wedge defined by Eqs.~\eqref{ccor} and \eqref{tcors}).
\begin{eqnarray}
B_\omega |_R (\tau,\zeta)& = & \frac{1}{\pi\sqrt{2}} e^{\frac{\omega\pi}{2}} e^{-i\omega\tau} K_{i\omega}(m\zeta) , \label{BmR} \\ 
B_\omega |_L (\tau,\zeta)& = & \frac{1}{\pi\sqrt{2}} e^{\frac{-\omega\pi}{2}} e^{i\omega\tau} K_{i\omega}(m\zeta) ,  \label{BmL} \\
B_\omega |_F (\tau,\zeta)& = & -\frac{i}{2^{3/2}} e^{\frac{\omega\pi}{2}} e^{-i\omega\tau} H_{i\omega}^{(2)} (m\zeta) ,  \label{BmF} \\
B_\omega |_P (\tau,\zeta)& = &  \frac{i}{2^{3/2}} e^{\frac{-\omega\pi}{2}} e^{-i\omega\tau} H_{i\omega}^{(1)} (m\zeta) ,  \label{BmP}
\end{eqnarray}
where $H_\nu^{(1),(2)}$ are Hankel functions. Note that in wedge $L$, $B_\omega (\tau,\zeta)$ has \emph{negative frequency}
\wrt{} $\tau$ because we have chosen the time translation generator  in $L$ to be $\tau^\mu = - b^\mu$, where
$b^\mu$ is the boost generator. 
Using Eqs.~(\ref{BmR}-\ref{BmP}) in Eqs.~\eqref{um} one obtais directly the representations of 
Unruh modes in each wedge. Modes $R_\omega$ are given by:
\begin{eqnarray}
R_\omega |_R (\tau,\zeta)& = &\frac{1}{\pi} \sqrt{\sinh{\pi\omega}} e^{-i\omega \tau} K_{i\omega}(m\zeta) , \label{errew} \\
R_\omega |_F (\tau,\zeta)& = &- \frac{i}{2^{3/2}} \frac{e^{-i\omega\tau}}{\sqrt{2\sinh(\pi\omega)}} \left[ e^{\omega\pi}H_{i\omega}^{(2)} (m\zeta) + e^{-\omega\pi} H_{i\omega}^{(1)} (m\zeta) \right] , \label{erreF}\\
R_\omega |_L(\tau,\zeta) & = & 0 , \\
R_\omega |_P(\tau,\zeta) & = & \frac{i}{ 2^{3/2}} \frac{ e^{-i\omega \tau}} { \sqrt{2\sinh{\pi\omega}}} \left[ H^{(1)}_{i\omega} (m\zeta) + H^{(2)}_{i\omega} (m\zeta) \right]  \,.
\end{eqnarray} 
And $L_\omega$  modes are
\begin{eqnarray}
L_\omega |_R (\tau,\zeta)& = & 0  , \\
L_\omega |_F (\tau,\zeta)& = &- \frac{i}{2^{3/2}} \frac{e^{i\omega\tau}}{\sqrt{2\sinh(\pi\omega)}} \left[ e^{\omega\pi}H_{i\omega}^{(2)} (m\zeta) + e^{-\omega\pi} H_{i\omega}^{(1)} (m\zeta) \right] , \label{eleF}\\
L_\omega |_L(\tau,\zeta) & = & \frac{1}{\pi} \sqrt{\sinh{\pi\omega}} e^{-i\omega \tau} K_{i\omega}(m\zeta) , \\
L_\omega |_P(\tau,\zeta) & = & \frac{i}{2^{3/2}} \frac{ e^{i\omega \tau}}{\sqrt{2\sinh{\pi\omega}}} \left[ H^{(1)}_{i\omega} (m\zeta) + H^{(2)}_{i\omega} (m\zeta) \right]  \,.
\end{eqnarray} 
Here we put some useful relations for the Hankel functions~\cite{erdelyi} from which one can simplify the expressions for Unruh modes
\beql{h1}
H^{(1)}_{\nu}(z)= \frac{1}{i\sin(\nu\pi)} \left[ J_{-\nu}(z)-e^{-i\nu\pi}J_\nu (z) \right] ,
\eeq
\beql{h2}
H^{(2)}_{\nu}(z)=  \frac{1}{i\sin(\nu\pi)} \left[ e^{i\nu\pi}J_\nu (z)-J_{-\nu}(z) \right]\,.
\eeq
These latter equations are used to obtain \equref{ImRF}. 

\section{Second order calculations}
\label{app:scnd}

Along this work we  do  several second order calculations of
expectation values of $\ffi (x) \ffi (x')$. In this Appendix we want to 
put the details of the calculus of $\tr\big(\ro_f^R \ffi_R (x) \ffi_R(x')\big)$ which leads to
\equref{ro2rr};
all the other calculi  are analogous to this one.

The state $\kets{f}$ takes the form of \equref{estgen} in the Unruh quantization,
in effect, from 
 \equref{ho} one can express   $\kets{f}$  as  (from now on we will omit the $\otimes$
in the Unruh states)  
\beql{fF}
\kets{f} = \sum_{J,K} F(J,K) \ketsl{J} \ketsr{K},
\qquad
F(J,K)=  F_0 (J,K) + F_1 (J,K) + F_2 (J,K) + \mathcal{O}(q^3) ,
\eeq
where the first term is $F_0 (J,K) = \mathcal{Q} F_{vac} (J,K)$ 
($\mathcal{Q}$ is defined above  \equref{ggaril}),
and 
  $F_{vac} (J,K)$ is the  
left-right superposition function defining  the inertial vacuum state $\km$ (\cf{} 
\equref{0mcorr}):
\beql{Fvac}
F_{vac} (J,K) =   Z^{1/2} e^{-\pi E(K)} \delta (J,K) ,
\eeq
where $Z$ is the normalization factor defined in \equref{Zeta}.
In order to express the other terms in \equref{fF}, it  is useful to
 define the normalization factor $N^\alpha_\th (K)$  by
\beql{Norm}
N^\alpha_\th (K) = \left\{
\begin{array}{ll}
\sqrt{K_\th +1} & \alpha=+\\
\sqrt{K_\th}    & \alpha=-
\end{array}\right. ,
\eeq
where  $K_\th$ is the particle component of state $\kets{K}_{L,R}$  in the mode with  frequency
centered on $\omega = \th$.
Using \equref{fii2} in \equref{ho} we find
\beqar
F_0 (J,K) & = & \mathcal{Q} Z^{1/2}e^{-\pi E(K)} \delta (J,K) \label{F0} , \\
F_1 (J,K) & = & -i  Z^{1/2} \Psi_\th  \, e^{-\pi E(J)}
\Big( N_\theta^+ (J) \delta (K,J+1_\theta ) 
+ N_\theta^- (J) \delta ( K, J-1_\theta ) \Big) \label{F1} , \\
F_2 (J,K) & = &  
-\frac{1}{2}  Z^{1/2} \Psi_\th^2  \, e^{-\pi E(J)} \Big( 
N_\theta^- (J) N_\theta^- (J-1_\theta )\; \delta (K,J-2_\theta ) \nonumber \\
& & \quad + 2 N_\theta^- (J) N_\theta^+ (J-1_\theta ) \;\delta (K,J ) 
+ N_\theta^+ (J) N_\theta^+ (J+1_\theta ) \;\delta (K,J+2_\theta ) \Big) \,.
\label{F2}
\eeqar
\equref{F0} is actually \equref{0mcorr} with the extra $\mathcal{Q}$ factor;  $E(J)$ is defined by \equref{e(j)}. 
The $\delta$ functions are defined by \equref{ladelta}. 
The  $J -1_\th$ which appears in the last term in
the \rhs{} of \equref{F1} corresponds to a normalized state $\ketsr{J-1_\th}$ defined by  
\beq
\hat{r}_\th \ketsr{J} = N^-_\th (J) \ketsr{J - 1_\th}
\eeq
with particle content $J - 1_\th = \{ J_{\omega_0}, \dots, J_\th -1,\dots\}$. The other terms
are defined analogously. 

The density matrix of state $\kets{f}$  reads
\beql{rf}
\hat{\rho}_f = \kets{f}\bras{f} =  \sum_{J,K,J',K'}  F(J,K) F(J',K')^*  \;  \ketsl{J}\ketsr{K}\;\brasl{J'}\brasr{K'} \,.
\eeq 
Let's write it 
in the following manner
\beql{rf2}
\hat{\rho}_f = \ro_f^{(0)} + \ro_f^{(1)} + \ro_f^{(2)} + \mathcal{O}(q^3) \,.
\eeq
Taking the trace over the left degrees of freedom to this expression we obtain that
\beql{ror2ap}
\ro_f^{R(2)} =   Z  \Psi_\th^2 \Big[-\frac{1}{2} ( e^{-2\pi\th}-1)^2 \ro_a -\frac{1}{2} ( e^{2\pi\th}-1)^2 \ro_b 
+ \ro_c\Big] ,
\eeq 
where
\beqar
\ro_a & =  & \sum_K e^{-2\pi E(K)} \sqrt{K_\th +2}\sqrt{K_\th +1} \: \ketsr{K}\brasr{K+2_\th} , \label{roa}\\
\ro_b & =  & \sum_K e^{-2\pi E(K)} \sqrt{K_\th}\sqrt{K_\th -1} \: \ketsr{K}\brasr{K-2_\th}, \label{rob}\\
\ro_c & = & \sum_K e^{-2\pi E(K)} \left(
-2K_\th + e^{2\pi\th} K_\th + e^{-2\pi\th}(K_\th+1) \right) \ketsr{K}\brasr{K}\,.\label{roc}
\eeqar
Writing the field operator as
\beql{fiap}
\ffi_R (x) = \sum_{\alpha=+,-} \sum_{m=0}^\infty R^{\alpha}_{\omega_m} (x) \hat{r}_{\omega_m}^\alpha
\eeq
where
\beq
R_{\omega_m}^+ (x) = R_{\omega_m}^* (x) , \qquad R_{\omega_m}^- (x) = R_{\omega_m} (x),
\qquad
\hat{r}_{\omega_m}^+  = \hat{r}_{\omega_m}^\dagger, \qquad \hat{r}_{\omega_m}^-  = \hat{r}_{\omega_m},
\eeq
we have 
\begin{multline}\label{roaff1ap}
\tr\big( \ro_a \ffi_R (x) \ffi_R (x')\big) = 
\sum_K 
\sum_{\alpha,\alpha'} \sum_{m,n=0}^\infty
e^{-2\pi E(K)} \sqrt{K_\th +2}\sqrt{K_\th +1} \:
N^{\alpha'}_{\omega_n} (K) N^\alpha_{\omega_m} (K + 1^{\alpha'}_{\omega_n}) \times \\
\times  R^{\alpha}_{\omega_m} (x) 
R^{\alpha'}_{\omega_n} (x') {}_R\!\bkt{K+2_\th}{ K + 1_{\omega_m}^\alpha + 1_{\omega_n}^{\alpha'}}_R  ,
\end{multline}
where $\ketsr{K+1^{\pm}_{\omega_m}} \equiv \ketsr{K\pm 1_{\omega_m}}$.
The expectation value appearing in the \rhs{} of \equref{roaff1ap} gives
\beql{hds}
{}_R\!\bkt{K+2_\th}{ K + 1_{\omega_m}^\alpha + 1_{\omega_n}^{\alpha'}}_R=
\delta_{\alpha,+}\;\delta_{\alpha',+}\;\delta_{{\omega_m},\th}\;\delta_{{\omega_n},\th}\,.
\eeq
From this one can read that  there is  only a  contribution to \equref{roaff1ap}  when the operator  $\ffi_R (x) \ffi_R (x')$ creates
two particles in the mode ${\omega}=\th$ in the \emph{state}  defined by $\ro_a$.
Using \equref{hds} in \equref{roaff1ap} we have that 
\beq\label{roaff1ap2}
\tr\big( \ro_a \ffi_R (x) \ffi_R (x')\big) = R_\th^* (x) 
R^{*}_{\th} (x') 
\sum_K  e^{-2\pi E(K)} (K_\th +2)(K_\th +1)  \,.
\eeq
To evaluate the sum in the \rhs{} of \equref{roaff1ap2} note that from \equref{suma} we have that
\beq\label{be10}\begin{split}
\sum_K  e^{-2\pi E(K)} f(K_{\th}) & = \sum_{K_{\th}=0}^\infty e^{-2\pi \th K_{\th}}f(K_{\th})\times
\prod_{
\substack{
m=0\\\omega_{m}\neq\th
}}^\infty
\sum_{K_{{\omega_m}}=0}^\infty
e^{-2\pi \omega_m K_{\omega_m}} \\
& =
\omom\, (1-e^{-2\pi \th}) \sum_{K_{\th}=0}^\infty e^{-2\pi \th K_{\th}}f(K_{\th}) ,
\end{split}\eeq
where $f(K_\th)$ is any function of $K_\th$ and  $\omom$ is given by \equref{0mnorm}.
For latter use it is convenient to define
\beql{Ge}
G_\th [f(K_\th) ]\equiv \sum_{K_{\th}=0}^\infty e^{-2\pi \th K_{\th}}f(K_{\th}) \,.
\eeq
Using \equref{be10} then 
\equref{roaff1ap2} becomes
\beql{trafin}
\tr\big( \ro_a \ffi_R (x) \ffi_R (x')\big) =   \omom \frac{2}{(1-e^{-2\pi\th})^2} R_\th^* (x) 
R^{*}_{\th} (x') ,
\eeq
where we have used that
\beq
G_\th [(K_\th +2)(K_\th +1)] = \frac{2}{(1-e^{-2\pi\th})^3} \,.
\eeq
Analogously we have
\beql{trbfin}
\tr\big( \ro_b \ffi_R (x) \ffi_R (x')\big) =   \omom \frac{2}{(e^{2\pi\th}-1)^2} R_\th (x) 
R_{\th} (x') \,.
\eeq
Before computing $\tr\big(\ro_c \ffi(x)\ffi(x')\big)$ let's define
\beql{hache}
H_{\omega_m} [K] (x,x') \equiv  K_{\omega_m} R_{\omega_m}^* (x) R_{\omega_m} (x') + (K_{\omega_m} +1)   R_{\omega_m} (x) R_{\omega_m}^* (x') \,.
\eeq
It can be verified that 
\beql{ins}
\tr\big(\ro_{vac}^R \ffi_R (x) \ffi_R (x')\big) = \sum_K e^{-2\pi E(K)}\sum_{m=0}^\infty H_{\omega_m} [K] (x,x') \,.
\eeq
Now we compute 
\begin{multline}\label{kand}
\tr\Big(\sum_K e^{-2\pi E(K)} \left[ e^{2\pi\th}K_\th + e^{-2\pi\th}(K_\th +1)\right]
\ketsr{K}\brasr{K}\ffi_R (x) \ffi_R (x')\Big) = \\
\begin{aligned}
= & \!\sum_K e^{-2\pi E(K)} \Big[ \big( e^{2\pi\th}K_\th^2 +  e^{-2\pi\th} K_\th (K_\th +1) \big) R_\th^* (x) R_\th (x') \\
  & \qquad \qquad   \qquad + \big( e^{2\pi\th}K_\th (K_\th +1) + e^{-2\pi\th} (K_\th +1)^2 \big) R_\th (x) R_\th^* (x') \Big]\\
  &  \qquad \qquad \qquad  \qquad   \qquad 
+ \sum_K e^{-2\pi E(K)} \big( e^{2\pi\th}K_\th +  e^{-2\pi\th}  (K_\th +1) \big) 
\sum_{\substack{
m=0\\\omega_{m}\neq\th
}}^\infty
 H_{\omega_m} [K] (x,x') \,.
\end{aligned}
\end{multline}
From the definition of $G_\th [f(K_\th)]$, \equref{Ge}, it can be shown that 
\beqar
e^{2\pi\th}G_\th [K_\th^2] + e^{-2\pi\th}G_\th [K_\th (K_\th +1)]  & = & G_\th [K_\th (2 K_\th + 1)] + G_\th [1] , \label{primis}\\
e^{2\pi\th}G_\th [K_\th (K_\th +1)] + e^{-2\pi\th}G_\th [(K_\th +1)^2]  & = & G_\th [(2K_\th +1) ( K_\th + 1)] + G_\th [1] , \label{seguns}\\
e^{2\pi\th}G_\th [K_\th ] + e^{-2\pi\th}G_\th [(K_\th +1)]  & = & G_\th [(2K_\th +1)] \,. \label{terces}
\eeqar
Now we use \equref{be10} to simplify the  \rhs{} of \equref{kand}. From Eqs.\eqref{primis}-\eqref{terces} we have that
\begin{multline}\label{kandin}
\tr\Big(\sum_K e^{-2\pi E(K)} \left[ e^{2\pi\th}K_\th + e^{-2\pi\th}(K_\th +1)\right]
\ketsr{K}\brasr{K}\ffi_R (x) \ffi_R (x')\Big) = \\
= \omom \big( R_\th^* (x) R_\th (x') + R_\th (x) R_\th^* (x') \big)  + 
\sum_K e^{-2\pi E(K)} \big( 2K_\th + 1 ) \big) \sum_{m=0}^\infty H_{\omega_m} [K] (x,x')
\,.\end{multline}
Then, from \equref{kandin} we have at once that (see \equref{roc})
\beql{troc}
\tr\big(\ro_c \ffi_R (x) \ffi_R (x') \big) = \omom \big( R_\th^* (x) R_\th (x') + R_\th (x) R_\th^* (x') \big)  + 
\tr\big(\ro_{vac}^R \ffi_R (x) \ffi_R (x')\big) ,
\eeq
where we have used \equref{ins}.
Finally, from Eqs.~\eqref{trafin},~\eqref{trbfin},~\eqref{troc} and~\eqref{ror2ap} we have
\beq\begin{split}
\tr\big(\ro_f^{R(2)}  \ffi_R (x) \ffi_R (x') \big) = & \Psi_\th^2\big( -R_\th (x) R_\th (x') - R_\th^* (x) R_\th^* (x') 
+ R_\th^* (x) R_\th (x') + R_\th (x) R_\th^* (x') \big) \\
& + Z  \Psi_\th^2 \tr\big(\ro_{vac}^R \ffi_R (x) \ffi_R (x')\big) 
\,.\end{split}\eeq
and \equref{ro2rr} follows directly.

\section{Inertial calculation}
\label{app:inertial}

In this Appendix we want to show that one obtains exactly \equref{DF} when computing the change in $\ev{\ffi(x)\ffi(x')}$ in the inertial scheme.
For this case we are going to use the scalar source given by \equref{jiner} written in the form
\beql{jinerap}
j(x) =  q \zeta_0 \frac{ \delta ( z - \sqrt{t^2 + \zeta_0^{2}})}{\sqrt{t^2 + \zeta_0^{2}}} ,
\eeq
where $\zeta_0 = 1/a$ and $a$ is the acceleration of the source.
Using the inertial field operator 
\beql{finer}
\ffi(x) = \int_{-\infty}^{\infty} dp \, \big( \psi_p (x) \hat{a}_p + \psi_p^* (x) \hat{a}^\dagger_p \big),
\eeq
where $\psi_p (x) $ represents a plane wave with frequency $\omega_p = +\sqrt{p^2 +m^2}$,
\beql{pws}
\psi_p (x) = \frac{1}{\sqrt{2\pi}\sqrt{2\omega_p}} e^{-i\omega_p t + i pz},
\eeq
in Eqs.~\eqref{hachei} and \eqref{efin},
it can be seen that the second order renormalized change in $\ev{\ffi(x)\ffi(x')}$ between states $\kets{f}$ and $\km$ is given by
\beql{cin}
C^{(2)}_{in} (x,x')= 4 \mathrm{Im} [Q(x)] \mathrm{Im} [Q(x')] ,
\eeq
where 
\beq
Q(x) = \int_M d^2 x'\, \int_{-\infty}^{\infty} dp\, j(x) \psi_p^* (x') \psi_p (x) \,.
\eeq 
Recall that for the quantized scalar field we are considering the positive frequency function is given by 
\beq
i\Delta^{(+)}(x,x')\equiv\bm \ffi(x)\ffi(x')\km = \int_{-\infty}^{\infty} dp\, \psi_p^* (x') \psi_p (x)
\eeq
and from~\cite{ruskis} we have the following result
\beql{frrus}	
\Delta^{(+)} (x,0) = \frac{1}{4}\times\left\{ 
\begin{array}{ll} 
H_{0}^{(2)} (m\sqrt{t^2 - z^2}) & t>\abs{z} \\
\frac{2i}{\pi} K_0 (m\sqrt{z^2 - t^2})  & \abs{t}<\abs{z} \\
-H_0^{(1)} (m\sqrt{t^2-z^2}) & t<-\abs{z} \\
\end{array} \right. \,.
\eeq
From now on we shall suppose that $x\in F$. Using the fact that $\Delta^{(+)} (x,x')=\Delta^{(+)} (x-x',0)$ and \equref{frrus} it can be seen that 
\beql{imq1}
\mathrm{Im} [Q(x)] = \frac{q\zeta_0}{4} \int_{-\infty}^{t^{-}} \frac{dt'}{\sqrt{t^{'\,2}+\zeta_0^2}} J_0 (m\sqrt{-2\sigma(t')}) ,
\eeq
where
\beq
\sigma (t') = \frac{1}{2} \left( - (t-t')^2 + \Big(z-\sqrt{t^{\prime\,2}+\zeta_0^2}\Big)^2\right)
\eeq
and $\sigma(t^{-}) =0$. To get \equref{imq1} we have used that $H^{(2)}_\nu (y)= J_\nu (y)-iY_\nu (y)$ and the fact that $J_0 (y)$, $Y_0 (y)$ are real for $y\geq 0$. 
Making the changes of variables $t' = \zeta_0 \sinh (\tau/\zeta_0 )$ and $u=\sqrt{-2\sigma(\tau)}$ we obtain
\beql{imq2}
\mathrm{Im}[Q(x)] = \frac{q\zeta_0}{2} \int_{0}^\infty \, 
\frac{u J_0 (m u)}{\sqrt{4 \zeta_0^2 \zeta^2 + (u^2 + \zeta_0^2 - \zeta^2)^2}}
\,du=\frac{q\zeta_0}{2} K_0 (m\zeta_0 ) J_0 (m\zeta ) ,
\eeq
were we have used $x=(t,z)$ and $t=\zeta\cosh(\tau_p /\zeta_0 )$,
$z=\zeta\sinh(\tau_p / \zeta_0 )$ (see \equref{cF}). 
The derivation that leads to the second equality in \equref{imq2}
is analogous to that  which leads to  Eq. \S 13.54(1) of \cite{watson}.
Thus we have proved that
\beq
C_{in}^{(2)} (x,x') = q^2 \zeta_0^2 K_0 (m\zeta_0)^2 J_0 (m\zeta)J_0 (m\zeta' ) ,
\eeq
which coincides functionally  with \equref{DF}.  In this computation one has to apply Wick theorem with the  notion of time and normal ordering associated to the inertial time parameter $t$.  However, the effect of this choice of time does not show up  in the physical change of the two point function but only in renormalization terms.
Note that, in contrast to the accelerated frame calculation, we did not need to introduce any regulator into the current (neither any cutoff as in the inertial frame calculation in~\cite{bremstralung2}).


\bibliography{biblioacc}

\begin{thebibliography}{18}
\expandafter\ifx\csname natexlab\endcsname\relax\def\natexlab#1{#1}\fi
\expandafter\ifx\csname bibnamefont\endcsname\relax
  \def\bibnamefont#1{#1}\fi
\expandafter\ifx\csname bibfnamefont\endcsname\relax
  \def\bibfnamefont#1{#1}\fi
\expandafter\ifx\csname citenamefont\endcsname\relax
  \def\citenamefont#1{#1}\fi
\expandafter\ifx\csname url\endcsname\relax
  \def\url#1{\texttt{#1}}\fi
\expandafter\ifx\csname urlprefix\endcsname\relax\def\urlprefix{URL }\fi
\providecommand{\bibinfo}[2]{#2}
\providecommand{\eprint}[2][]{\url{#2}}

\bibitem[{\citenamefont{Boulware}(1980)}]{boulware}
\bibinfo{author}{\bibfnamefont{D.~G.} \bibnamefont{Boulware}},
  \bibinfo{journal}{Ann. Phys. (N.Y.)} \textbf{\bibinfo{volume}{124}},
  \bibinfo{pages}{169} (\bibinfo{year}{1980}).

\bibitem[{\citenamefont{Unruh}(1976)}]{efectounruh}
\bibinfo{author}{\bibfnamefont{W.~G.} \bibnamefont{Unruh}},
  \bibinfo{journal}{Phys. Rev.} \textbf{\bibinfo{volume}{D14}},
  \bibinfo{pages}{870} (\bibinfo{year}{1976}).

\bibitem[{\citenamefont{Unruh and Wald}(1984)}]{waldunruh}
\bibinfo{author}{\bibfnamefont{W.~G.} \bibnamefont{Unruh}} \bibnamefont{and}
  \bibinfo{author}{\bibfnamefont{R.~M.} \bibnamefont{Wald}},
  \bibinfo{journal}{Phys. Rev.} \textbf{\bibinfo{volume}{D29}},
  \bibinfo{pages}{1047} (\bibinfo{year}{1984}).

\bibitem[{\citenamefont{Higuchi
  et~al.}(1992{\natexlab{a}})\citenamefont{Higuchi, Matsas, and
  Sudarsky}}]{bremstralung}
\bibinfo{author}{\bibfnamefont{A.}~\bibnamefont{Higuchi}},
  \bibinfo{author}{\bibfnamefont{G.~E.~A.} \bibnamefont{Matsas}},
  \bibnamefont{and} \bibinfo{author}{\bibfnamefont{D.}~\bibnamefont{Sudarsky}},
  \bibinfo{journal}{Phys. Rev.} \textbf{\bibinfo{volume}{D45}},
  \bibinfo{pages}{R3308} (\bibinfo{year}{1992}{\natexlab{a}}).

\bibitem[{\citenamefont{Wald}(1985)}]{librowald}
\bibinfo{author}{\bibfnamefont{R.~M.} \bibnamefont{Wald}},
  \emph{\bibinfo{title}{Quantum field theory in curved space-time and black
  hole thermodynamics}} (\bibinfo{publisher}{University Press},
  \bibinfo{address}{Chicago}, \bibinfo{year}{1985}).

\bibitem[{\citenamefont{Narozhny et~al.}(2002)\citenamefont{Narozhny, Fedotov,
  Karnakov, Mur, and Belinskii}}]{ruskis}
\bibinfo{author}{\bibfnamefont{N.~B.} \bibnamefont{Narozhny}},
  \bibinfo{author}{\bibfnamefont{A.~M.} \bibnamefont{Fedotov}},
  \bibinfo{author}{\bibfnamefont{B.~M.} \bibnamefont{Karnakov}},
  \bibinfo{author}{\bibfnamefont{V.~D.} \bibnamefont{Mur}}, \bibnamefont{and}
  \bibinfo{author}{\bibfnamefont{V.~A.} \bibnamefont{Belinskii}},
  \bibinfo{journal}{Phys. Rev.} \textbf{\bibinfo{volume}{D65}},
  \bibinfo{pages}{025004} (\bibinfo{year}{2002}).

\bibitem[{\citenamefont{Fulling and Unruh}(2004)}]{fullvsruskis}
\bibinfo{author}{\bibfnamefont{S.~A.} \bibnamefont{Fulling}} \bibnamefont{and}
  \bibinfo{author}{\bibfnamefont{W.~G.} \bibnamefont{Unruh}},
  \bibinfo{journal}{Phys. Rev.} \textbf{\bibinfo{volume}{D70}},
  \bibinfo{pages}{048701} (\bibinfo{year}{2004}).

\bibitem[{\citenamefont{Kay}(1978)}]{kay}
\bibinfo{author}{\bibfnamefont{B.~S.} \bibnamefont{Kay}},
  \bibinfo{journal}{Commun. Math. Phys.} \textbf{\bibinfo{volume}{62}},
  \bibinfo{pages}{55} (\bibinfo{year}{1978}).

\bibitem[{\citenamefont{Ashtekar and Magnon}(1975)}]{ashtekar}
\bibinfo{author}{\bibfnamefont{A.}~\bibnamefont{Ashtekar}} \bibnamefont{and}
  \bibinfo{author}{\bibfnamefont{A.}~\bibnamefont{Magnon}},
  \bibinfo{journal}{Proc. Roy. Soc. Lond.} \textbf{\bibinfo{volume}{A346}},
  \bibinfo{pages}{375} (\bibinfo{year}{1975}).

\bibitem[{\citenamefont{Fulling}(1973)}]{fulling}
\bibinfo{author}{\bibfnamefont{S.~A.} \bibnamefont{Fulling}},
  \bibinfo{journal}{Phys. Rev.} \textbf{\bibinfo{volume}{D7}},
  \bibinfo{pages}{2850} (\bibinfo{year}{1973}).

\bibitem[{\citenamefont{Watson}(1922)}]{watson}
\bibinfo{author}{\bibfnamefont{G.~N.~A.} \bibnamefont{Watson}},
  \emph{\bibinfo{title}{A Treatise on the Theory of {B}essel Functions}}
  (\bibinfo{publisher}{The {U}niversity {P}ress}, \bibinfo{address}{Cambridge,
  UK}, \bibinfo{year}{1922}).

\bibitem[{\citenamefont{Kay}(1985)}]{kaydouble}
\bibinfo{author}{\bibfnamefont{B.~S.} \bibnamefont{Kay}},
  \bibinfo{journal}{Commun. Math. Phys.} \textbf{\bibinfo{volume}{100}},
  \bibinfo{pages}{57} (\bibinfo{year}{1985}).

\bibitem[{\citenamefont{Gerlach}(1988)}]{gerlach}
\bibinfo{author}{\bibfnamefont{U.~H.} \bibnamefont{Gerlach}},
  \bibinfo{journal}{Phys. Rev.} \textbf{\bibinfo{volume}{D38}},
  \bibinfo{pages}{514} (\bibinfo{year}{1988}).

\bibitem[{\citenamefont{Martinetti and Rovelli}(2003)}]{rovelli}
\bibinfo{author}{\bibfnamefont{P.}~\bibnamefont{Martinetti}} \bibnamefont{and}
  \bibinfo{author}{\bibfnamefont{C.}~\bibnamefont{Rovelli}},
  \bibinfo{journal}{Class. Quant. Grav.} \textbf{\bibinfo{volume}{20}},
  \bibinfo{pages}{4919} (\bibinfo{year}{2003}).

\bibitem[{\citenamefont{Horodecki et~al.}(2001)\citenamefont{Horodecki,
  Horodecki, and Horodecki}}]{horodeckis1}
\bibinfo{author}{\bibfnamefont{M.}~\bibnamefont{Horodecki}},
  \bibinfo{author}{\bibfnamefont{P.}~\bibnamefont{Horodecki}},
  \bibnamefont{and}
  \bibinfo{author}{\bibfnamefont{R.}~\bibnamefont{Horodecki}},
  \emph{\bibinfo{title}{in: Quantum Information: An Introduction to Basic
  Theoretical Concepts and Experiments}} (\bibinfo{publisher}{Springer-Verlag},
  \bibinfo{year}{2001}), Springer Tracts in Modern Physics, 173.

\bibitem[{\citenamefont{Higuchi
  et~al.}(1992{\natexlab{b}})\citenamefont{Higuchi, Matsas, and
  Sudarsky}}]{bremstralung2}
\bibinfo{author}{\bibfnamefont{A.}~\bibnamefont{Higuchi}},
  \bibinfo{author}{\bibfnamefont{G.~E.~A.} \bibnamefont{Matsas}},
  \bibnamefont{and} \bibinfo{author}{\bibfnamefont{D.}~\bibnamefont{Sudarsky}},
  \bibinfo{journal}{Phys. Rev.} \textbf{\bibinfo{volume}{D46}},
  \bibinfo{pages}{3450} (\bibinfo{year}{1992}{\natexlab{b}}).

\bibitem[{\citenamefont{Boulware}(1975)}]{boulware_qntzn}
\bibinfo{author}{\bibfnamefont{D.~G.} \bibnamefont{Boulware}},
  \bibinfo{journal}{Phys. Rev.} \textbf{\bibinfo{volume}{D11}},
  \bibinfo{pages}{1404} (\bibinfo{year}{1975}).

\bibitem[{\citenamefont{Erd\'elyi}(1953)}]{erdelyi}
\bibinfo{author}{\bibfnamefont{A.}~\bibnamefont{Erd\'elyi}},
  \emph{\bibinfo{title}{Higher trascendental functions}},
  vol.~\bibinfo{volume}{II} (\bibinfo{publisher}{Robert E.~Krieger},
  \bibinfo{address}{USA}, \bibinfo{year}{1953}).

\end{thebibliography}
\end{document}